\documentclass[11pt,a4paper]{article}
\usepackage{jheppub}
\usepackage[T1]{fontenc}
\title{Observables for General Relativity related to geometry}

\author[a]{Pawe{\l} Duch,}
\author[b]{Wojciech Kami\'nski,}
\author[b,c]{Jerzy Lewandowski}
\author[b]{and J\k{e}drzej \'Swie\.zewski}

\affiliation[a]{Institute of Physics, Jagiellonian University,\\ Reymonta 4, 30-059 Krak\'{o}w, Poland} 
\affiliation[b]{Faculty of Physics, University of Warsaw,\\ Ho\.{z}a 69, 00-681 Warszawa, Poland}
\affiliation[c]{Institute for Quantum Gravity (IQG), FAU Erlangen -- Nurnberg,\\ Staudtstr. 7, 91058 Erlangen, Germany}

\emailAdd{pawel.duch@uj.edu.pl} 
\emailAdd{wojciech.kaminski@fuw.edu.pl} 
\emailAdd{jerzy.lewandowski@fuw.edu.pl}
\emailAdd{swiezew@fuw.edu.pl}

\abstract{
We present a new scheme of defining invariant observables for general relativistic systems. The scheme is based on the introduction of an observer which endowes the construction with a straightforward physical interpretation. The observables are invariant with respect to spatial diffeomorphisms which preserve the observer. The limited residual spatial gauge freedom is studied and fully understood. A full canonical analysis of the observables is presented: we analyze their variations, Poisson algebra and discuss their dynamics. Lastly, the observables are used to solve the vector constraint, which triggers a possible considerable reduction of the degrees of freedom of general relativistic theories.
}
      

\begin{document}
\maketitle
\flushbottom

\section{Introduction}\label{Introduction}

\subsection{Relate observables to spacetime geometry}

A spacetime point is not a physical notion according to the dynamics of General Relativity \cite{Einstein}. Moreover, in General Relativity, a manifold is what replaces Euclidean space to be the mathematical structure best suited to be endowed with geometry and turned into a spacetime. This is the case as long as we consider particles and fields interacting with external, background spacetimes. However, when spacetimes are enlived by the dynamics, then the status of the manifold and its points changes. Given a physical general relativistic system described by fields  $\phi_1,\ldots,\phi_n$ (including the gravitational field), it is only the diffeomorphism invariant information that has a physical meaning. For example, we can distinguish any $4$ of the (real valued components of the) fields, say  $\phi_1,\ldots,\phi_4$, such that every remaining field $\phi_{4+\alpha}$ defines a function 
\[       
\phi_{4+\alpha}\ =\ {\Phi}_\alpha(\phi_1,\ldots,\phi_4).
\]
This method is called deparametrization. 

In the study of geometry of curved spacetimes, the problem of the diffeomorphism invariant characterization has been solved long time ago by \'{E}lie Cartan as ``the equivalence problem'' (when are two geometries difffeomorphism-equivalent?) \cite{Cartan}. Cartan's solution relies on an invariant system of coordinates consisting of scalars constructed from the Riemann tensor and its covariant derivatives. The system becomes degenerate for nongeneric geometries, but that is pardonable. The problem is, that the system is totally blind in a flat, Minkowski spacetime. Another weak point of this system is that matter seems not to be coupled to the derivatives of curvature. It seems more natural to expect that matter is sensitive to distances.  

Another example of a deparametrization of spacetime is the GPS construction \cite{RovelliGPS}. Notice, however, that as geometric as it is, GPS needs at least $n$ observers to identify points of an $n$-dimensional spacetime. It is worth noting, that there are modifications of the idea of GPS which use only two observers to define a sensible observable \cite{Khavkine, KhavkineBonga}.

Another approach to the problem of the diffeomorphism invariant descriptions of general relativistic systems is a deparametrization of the theory performed entirely using a material medium: for example elastic \cite{Kiju} or dust \cite{BrownKuchar}.  A field theoretical version of the deparametrisation uses scalar fields \cite{RovelliSmolin}. 

The requirement of the diffeomorphism invariance in the description of general relativistic systems is reflected by an emergence of first class constraints and gauge transformations in the canonical formulation of the theory. Physical observables in this framework have to be gauge invariant, and we call them Dirac observables.    

In this work we propose a deparametrization of a general relativistic theory with the distances and angles defined by the geometry and by a single observer. We derive the corresponding canonical framework. It leads naturally to a construction of Dirac observables. We study their properties. Technically, this is the most difficult part of our task.

\subsection{Fields and framework}\label{sectionFandF}

We consider canonical gravity coupled to some matter fields. The theory consists of a 3-manifold $\Sigma$ and canonically conjugate pairs of fields defined thereon: a 3-metric tensor $q_{ij}$ and its momentum $p^{ij}$, remaining fields $\phi_\alpha$ and $\pi^\alpha$, $\alpha=1,\ldots,n$. Each set of fields $(q,p,\phi_\alpha,\pi^\alpha)$ is a point in the kinematical phase space $\Gamma$ of the considered theory. The Poisson brackets are 
\begin{equation}\label{ccr}
\{q_{ij}(\sigma),\ p^{kl}(\sigma')\}\ =\ \delta^k_{(i}\delta^l_{j)}\delta(\sigma,\sigma'),\qquad \{\phi_\alpha(\sigma),\ \pi^{\alpha'}(\sigma')\}\ =\ \delta_\alpha^{\alpha'}\delta(\sigma,\sigma'), 
\end{equation}
 
In Section \ref{sekcjarozwdiff}, physical points of $\Gamma$ are selected, as those which satisfy the Arnowitt-Deser-Misner (ADM \cite{ADM}) vector constraints ${C}_i(\sigma)=0$ which in the spacetime approach generate the diffeomorphisms of the Cauchy surface, and in the phase space $\Gamma$ generate the induced action of the diffeomorphisms of $\Sigma$.

\section{Deparametrization by 3-geometry}

\subsection{Idea}\label{Idea}

For every $(q,p,\phi_\alpha,\pi^\alpha)\in\Gamma$ we want to use the 3-geometry $q$ to characterize each point  $\sigma\in \Sigma$ by: \emph{(i)} the geodesic distance to an observer, \emph{(ii)} the point at the observers sphere of directions corresponding to the geodesic curve passing through $\sigma$ which reaches the observer. That characterization is $q$-dependent. Expressing all the fields $q,\ldots,\pi^\alpha$ defined on $\Sigma$ in terms of the coordinates just described results in functions on $\Gamma$ {\it invariant} with respect to those diffeomorphisms which act trivially on the observer. The remaining scalar constraint will be deparametrized by one of the fields $\phi_\alpha$ in this approach.

\subsection{Observers description of 3-geometry -- adapted coordinates}

Technically speaking what we do is fix a point $\sigma_0\in\Sigma$ and a frame $e^0_I\in T_{\sigma_0}\Sigma$, where $I = 1,2,3$. Working in a point $(q,p,\phi_\alpha,\pi^\alpha)\in\Gamma$ we introduce an orthonormal frame $e_I$ obtained from the fixed $e^0_I$ by a Gram-Schmidt orthonormalisation process, namely
\begin{equation}
e_I\ =\ \overset{3}{\underset{J=1}{\sum}}M_{IJ}e^0_J,
\end{equation}
where the matrix $M$ is unique since the process requires it to be lower-triangular and have positive entries on the diagonal.\footnote{Alternatively, we could say that we fix a family of frames, related to each other by transformations with lower-triangular matrices having positive entries on the diagonal. Then for each metric, there exists exactly one frame from our family which is orthonormal with respect to that metric. This frame can be found by picking any of the members and performing on it the Gram-Schmidt process involving the given metric.} Of course, both the matrix $M$ and the orthonormal frame $e_I$ are $q$-dependent. To every point $\sigma$ in a neighborhood of $\sigma_0$ we assign three numbers which we will collectively denote as $x$ (envoking specific one of them we will use an index, e.g. $x^I$ is the $I$-th from the three) such that
\begin{equation}
{\rm exp}_{\sigma_0}(x^I e_I)\ =\ \sigma,
\end{equation}
where the ${\rm exp_{\sigma_0}(\cdot)}$ map is the exponent map sending vectors from $T_{\sigma_0}\Sigma$ to points in the manifold $\Sigma$. We will refer to the coordinates $(x^I)$ as the \emph{Cartesian coordinates adapted to $q$}.\footnote{In different contexts the presented coordinates are often called Riemann normal coordinates. We choose a different name here to underline that we have different sets of coordinates for different phase space points.}

Throughout the paper we will also use coordinates related to the Cartesian adapted coordinates by a simple (phase-space independent) renaming. For every point $\sigma$ to which we assigned coordinates $(x^I)$, we assign three numbers $(r,\theta)$ (where $\theta$ collectively denotes two values; they will be further referred to with the use of an index $A$ assuming two values) such that
\begin{equation}\label{relationcartspher}
x^I\ =\ rn^I(\theta),
\end{equation}
where $n^I(\theta)$ is a unit vector in $T_{\sigma_0}\Sigma$, so that the two angular parameters $\theta$ parametrize surfaces of constant radial distance $r$ from the observer. We will denote those coordinates $(y^a)(\sigma)=(y^r,y^A)(\sigma)=(r,\theta)$ and call them \emph{spherical coordinates adapted to $q$}.

A few remarks concerning the adapted coordinates follow:
\begin{itemize}
\item The spherical adapted coordinates have a very natural physical interpretation. If we treat the point $\sigma_0$ and the frame $e_I$ to be an observers position and his way of parametrising spatial directions respectively, then the value of $r$ represents the proper distance between the observer and a given point in his neighborhood and $\theta$ denotes the angles at which the point is located with respect to his directions.

\item Both sets of the adapted coordinates we introduced above paramterize a neighborhood of the observer as long as the ${\rm exp}$ map is injective. The size of such a neighborhood depends on the metric $q$, but one can show that there always exists a neighborhood of the observer in which the coordinates are well-defined. Note also that only the Cartesian adapted coordinates are regular at the point $\sigma_0$, since the spherical coordinates are subject to the usual limitations of standard spherical coordinates in this point.

\item The maps
\begin{equation}
q \mapsto (y^a)\qquad{\rm and}\qquad q\mapsto (x^I)
\end{equation}
are invariant with respect to those diffeomorphisms $\psi:\Sigma\rightarrow \Sigma$ for which
\begin{equation}
\psi(\sigma_0) = \sigma_0\qquad{\rm and}\qquad \psi'(\sigma_0) \ =\ M,
\end{equation}  
where $M$, when expressed in the frame $e^0_I$, is lower-triangular and has positive entries on the diagonal. Let us denote the subgroup of such diffeomorphisms by Diff$_{\text{obs}}$. 

\item An expansion in $r$ of a metric $q$ expressed in the coordinates adapted to it is given by
\begin{equation}\label{rozwinieciewzerze}
q_{IJ}(x)\ =\ \delta_{IJ} + {\cal O}(r^2)\qquad{\rm and}\qquad q_{AB}(r,\theta)\ =\ r^2\eta_{AB}(\theta) + {\cal O}(r^4).
\end{equation}

\item An interesting question to ask is when, given a metric tensor and a coordinate system in a neighborhood of $\sigma_0$, the given coordinates are the spherical coordinates adapted to the given metric. Let $q$ be a metric tensor in $\Sigma$ and suppose $(y^a)$ are 
the spherical coordinates adapted to $q$. Let $q'$ be another metric tensor. Suppose that in terms of the coordinates adapted to $q$ 
\begin{equation}
q'\ =\ dy^r\otimes dy^r + q'_{AB}dy^A\otimes dy^B,
\end{equation}
in an open neighborhood ${\cal U}$ of the point $\sigma_0$ such that ${\cal U}$ intersects each curve
\begin{equation}
y^A\ =\ {\rm const}
\end{equation}
on a connected segment. Then, the coordinates $(y'^a)$ adapted to $q'$ coincide with $(y^a)$ on ${\cal U}$  
\begin{equation}
\left.(y'^a)\right|_{\cal U}\ =\ \left.(y^a)\right|_{\cal U}. 
\end{equation}
Indeed, for both metric tensors the curves tangent to $\delta_r$ are geodesic, $r$ measures the distance along them from $\sigma_0$, and necessarily
\begin{equation}
q(\sigma_0)\ =\ q'(\sigma_0).
\end{equation}
\end{itemize}

\subsection{Diff$_{\text{obs}}$-invariant observables in adapted coordinates}\label{invobsinadaptedcoordinates}

Given a point $(q,p,\phi_\alpha,\pi^\alpha)\in\Gamma$ and values $(r,\theta)$ assumed by the adapted spherical coordinates 
at a point $\sigma\in\Sigma$, we can introduce the following functions on $\Gamma$:
\begin{subequations}\label{invobs}
\begin{align}
Q_{ab}(r,\theta)&:(q,p,\phi_\alpha,\pi^\alpha)\ \mapsto\ q_{ab}(r,\theta),\\
P^{ab}(r,\theta)&:(q,p,\phi_\alpha,\pi^\alpha)\ \mapsto\ p^{ab}(r,\theta),\label{Pirtheta}\\
\Phi_\alpha(r,\theta)&:(q,p,\phi_\alpha,\pi^\alpha)\ \mapsto\ \phi_\alpha(r,\theta),\\
\Pi^\alpha(r,\theta)&:(q,p,\phi_\alpha,\pi^\alpha)\ \mapsto\ \pi^\alpha(r,\theta),
\end{align}
\end{subequations}
where by
\begin{equation}
q_{ab}(r,\theta),\ p^{ab}(r,\theta),\ \phi_\alpha(r,\theta),\ \pi^\alpha(r,\theta) 
\end{equation}
we mean the components of the metric $q$, the conjugate momentum $p$, the values of the fields $\phi_\alpha$ and $\pi^\alpha$ in the spherical coordinates adapted to $q$, evaluated at the point $\sigma$ at which the adapted coordinates assume the values $(r,\theta)$.
The domain of the above functions consists of those points $(q,p,\phi_\alpha,\pi^\alpha)$ of $\Gamma$, for which a given triple of the values $(r,\theta)$ is in the range of the coordinates adapted to $q$. 

The functions (\ref{invobs}) have a very important property: each of them is {\it invariant} with respect to the action in the phase space 
$\Gamma$ of every element of the subgroup Diff$_{\text{obs}}$ defined above. The connected component of the identity of that subgroup is generated by the vector constraints $C(\vec{N})\ =\ \int d^3 \sigma N^i (\sigma)C_i(\sigma)$, such that 
\begin{equation}\label{diffgen}
N^I(\sigma_0)\ =\ 0, \qquad{\rm and}\qquad N^I_{,J}(\sigma_0) \text{ is\ lower-triangular} 
\end{equation}
in the frame $e^0_I$ or, equivalently, in the frame $e_I$.

Moreover, some of the invariant functions  $Q_{ab}(r,\theta)$ are trivial. Indeed, 
\begin{equation}\label{warunkigauge}
Q_{rr}(r,\theta)\ =\ 1,\qquad  Q_{rA}(r,\theta)\ =\ 0
\end{equation}
identically by construction.

Therefore, 
\begin{equation}\label{obs}
F(r,\theta)\in\{Q_{AB}(r,\theta),\ P^{AB}(r,\theta),\ P^{rr}(r,\theta),\ P^{rA}(r,\theta),\ \Phi_\alpha(r,\theta),\ \Pi^\alpha(r,\theta)\},
\end{equation} 
represent a complete set of the Diff$_{\text{obs}}$-invariant degrees of freedom of geometry and fields near the point $\sigma_0$
as $(r,\theta)$ range through all the values assumed by the spherical coordinates $(y^a)$.

{\vspace{0.5cm}}

In an analogous way, we use the adapted Cartesian coordinates to define observables $G(x)$, where $G\in\{Q_{IJ}, P^{IJ},
\Phi_\alpha,\Pi^\alpha\}$. The relation between the observables defined by the adapted spherical and, respectively, Cartesian coordinates is the usual transformation
\begin{subequations}\label{cyltocart}
\begin{align}
Q_{ab}(r,\theta)\ &=\ \frac{\partial x^I}{\partial y^a}\frac{\partial x^J}{\partial y^b}Q_{IJ}(x), \\
P^{ab}(r,\theta)\ &=\ \left|\frac{\partial(x^1,x^2,x^3)}{\partial (r,\theta^1,\theta^2)}\right|\frac{\partial y^a}{\partial x^J}\frac{\partial y^b}{\partial x^J}P^{IJ}(x)
\end{align}
\end{subequations}
and similarly for the matter fields $\Phi_\alpha,\Pi^\alpha$, where the labels $(r,\theta)$ and $(x)$ are such that they point at the same point $\sigma$ in coordinates adapted to the same metric. In the above expressions we have used the transformation 
\begin{equation}\label{xtorn}
x^I(r,\theta)\ =\ rn^I(\theta)
\end{equation}
and its inverse. Notice, that in this transformation the Jacobi matrices play the role of the labels only. This happens because the relation between the Cartesian and spherical adapted coordinates does not depend on the phase space point, as the adapted coordinates themselves do. 

An advantage of the observables $G(x)$ is that they extend in a regular way to the point $\sigma_0$ itself:
\begin{equation}\label{obsx_0}
P^{IJ}(0), \Phi_\alpha(0), \Pi^\alpha(0).
\end{equation}
However, due to (\ref{rozwinieciewzerze}) 
\begin{equation}
Q_{IJ}(0)\ =\ \delta^{IJ}
\end{equation}
by construction.

{\vspace{0.5cm}}

In Section \ref{sekcjarozwdiff}, we will consider the subspace $\Gamma_{C}\subset\Gamma$ defined by the vanishing of the constraints (\ref{diffgen}). Thereon, the functions $P^{rr}(r,\theta)$ and $P^{rA}(r,\theta)$ will be determined by the remaining functions $Q_{AB}(r,\theta),\ P^{AB}(r,\theta),\ \Phi_\alpha(r,\theta),\ \Pi^\alpha(r,\theta)$. Which means that in $\Gamma_{C}$ a smaller set of independent observables can be indentified.

\subsection{Observables in a general coordinate system}

At a point $(q,p,\phi_\alpha,\pi^\alpha)\in \Gamma$, the relation between the values of the invariant observables (\ref{obs}) on the left hand side and the values of the corresponding fields expressed in some general coordinates $(z^i)$ is given by
\begin{subequations}\label{Obsexpl}
\begin{align}
\Phi_\alpha(r,\theta)\ &=\ \rho\left(\frac{\partial y}{\partial z}\right)^\beta_\alpha\phi_\beta(\sigma^q_{(r,\theta)}),\\
\Pi^\alpha(r,\theta)\ &=\ \left|\det \left(\frac{\partial z}{\partial y}\right)\right|\rho^*\left(\frac{\partial y}{\partial z}\right)_\beta^\alpha\pi^\beta(\sigma^q_{(r,\theta)}),\\
Q_{AB}(r,\theta)\ &=\ \frac{\partial z^i}{\partial y^A}\frac{\partial z^j}{\partial y^B}q_{ij}(\sigma^q_{(r,\theta)}),\\
P^{AB}(r,\theta)\ &=\ \left|\det \left(\frac{\partial z}{\partial y}\right)\right|\frac{\partial y^A}{\partial z^i}\frac{\partial y^B}{\partial z^j} p^{ij}(\sigma^q_{(r,\theta)}),
\end{align}
\end{subequations}
where we are taking into account a general case when the fields $\phi_\alpha$ are not just scalar and the index $\alpha$ transforms with changes of coordinates according to a representation $\rho$ and the point $\sigma^q_{(r,\theta)}$ at which all the fields on the right hand sides are evaluated is a function of the phase space point $(q,p,\phi_\alpha,\pi^\alpha)$ determined by the condition
\begin{equation}
(y^a(\sigma^q_{(r,\theta)}))\ =\ (r,\theta).
\end{equation}
From (\ref{Obsexpl}) we see clearly the dependence of the observables on the canonical data. Note, that the observables depend on the field they are constructed from directly, and additionally, they have a very nontrivial dependance on the metric through the usage of coordinates adapted to it. Needless to say, a relation analogous to (\ref{Obsexpl}) for the observables $G(x)$ can easily be spelled out. We will write the observables in one more way, that will be useful, for example, for calculating their Poisson brackets.

\section{Variations of the observables in terms of adapted coordinates}

Although our definition of observables (\ref{invobs}) may be applied to matter fields of arbitrary type, the derivation of the variations and Poisson brackets of those matter observables depends on that type. Therefore, in the following we limit ourselves to the case of matter fields which are scalar, meaning that from now on, for each $\alpha$, $\phi_\alpha$ is a scalar field and $\pi^\alpha$ is a scalar density.

As has been already noted above, from the point of view of the phase space dependance the observables $G(x)$ and $F(r,\theta)$ differ only by a relabelling. Hence it is enough to study the variations of the observables $G(x)$ to be able to determine the variations of the observables $F(r,\theta)$ as well. Therefore, in the rest of this section we concentrate on the observables $G(x)$.

Calculating the Poisson brackets between the observables $G(x)$ and any other functions defined on $\Gamma$ involves the functional derivatives,
\begin{subequations}
\begin{align}
\{G(x),\ p^{ij}(\sigma)\}\ &=\ \frac{\delta G(x)}{\delta q_{ij}(\sigma)},\\
\{G(x),\  q_{ij}(\sigma)\}\ &=\ -\frac{\delta G(x)}{\delta p^{ij}(\sigma)},\\
\{G(x),\ \pi^\alpha(\sigma)\}\ &=\ \frac{\delta G(x)}{\delta \phi_\alpha(\sigma)},\\
\{G(x),\ \phi_\alpha(\sigma)\}\ &=\ -\frac{\delta G(x)}{\delta \pi^\alpha(\sigma)}.
\end{align}
\end{subequations}
The functional derivatives can be calculated by variating (\ref{Obsexpl}) directly (and using the relations (\ref{cyltocart})). This however, is complicated since it involves also the variation of a solution of the geodesic equation $\sigma^q_{(x)}$ with respect to the metric $q$. However, many properties of those variations can be deduced in an easier way. In fact, eventually, in the way we will present below the functional derivatives of the observables $G(x)$ (and in consequence also $F(r,\theta)$) can be determined completely. 

Throughout this section, given an observable $G(x)$, and a point $(\check q,\check p,\check \phi_\alpha,\check \pi^\alpha)\in \Gamma$ we fix the coordinates $(\check x^I)$ adapted to $\check q$. We study the variations
\begin{equation}
\left.\frac{d}{d\epsilon}\right|_{\epsilon=0}\left.G(x)\right|_{(\check q + \epsilon\delta q, \check p + \epsilon\delta p, \check \phi_\alpha+\epsilon\delta \phi_\alpha, \check \pi^\alpha+\epsilon\delta \pi^\alpha)}
\end{equation}
and express them in the coordinates $(\check x^I)$. It will also be convenient to use the spherical coordinates $(\check y^a)$ adapted to $\check q$.

A special role will be played by the geodesic line consisting of the points
\begin{equation}
\sigma^{\check q}_{(r',\theta)}\quad\text{ for }0< r'\leq r,
\end{equation}
which is a unique geodesic line connecting the points $\sigma_0$ and $\sigma^{\check q}_{(x)}$ if only the labels $(r,\theta)$ and $(x)$ are such that they indicate the same point in respective coordinates adapted to the metric $\check q$.

\subsection{Yet another formula for the observables}\label{yetanotherformula}

In this section we will explicitely discuss the observables defined with the use of the Cartesian adapted coordinates $G(x)$, but all the arguments and results obtained below apply also in the case of the observables $F(r,\theta)$.

\vspace{0.5cm}

Given an observable $G(x)$ and a phase space point  $\check\gamma=(\check q,\check p,\check \phi_\alpha,\check \pi^\alpha)$ at which we want to calculate the variations, the idea is to decompose the observable into two parts: a part which will just be the appropriate field corresponding to $G$ (we will denote it by $g$) expressed in coordinates $(\check x^I)$ adapted to $\check q$ and the difference $D$. That is  
\begin{equation}\label{obsdecomp} 
\left.G(x)\right|_{(q,p,\phi_\alpha,\pi^\alpha)}\ =\ \left.g(x)\right|_{(q,p,\phi_\alpha,\pi^\alpha)}+\left.D_{G(x)}\right|_{(q,p,\phi_\alpha,\pi^\alpha)},
\end{equation} 
where the indices $I,J,\alpha$ of the field (hidden in the symbol $g$) correspond to the fixed coordinates $(\check x^I)$ (notice, that on the left hand side, according to our earlier definitions, the indices $I,J,\alpha$ correspond to the coordinates $(x^I)$ adapted to the current metric $q$). Note, that the points of $\Sigma$ at which the fields in $G(x)$ and $g(x)$ are evaluated in general differ, since the first one is the point $\sigma^q_{(x)}$ and the second one is $\sigma^{\check q}_{(x)}$.

\begin{figure}[h]
\centering
\includegraphics[width=0.6\textwidth]{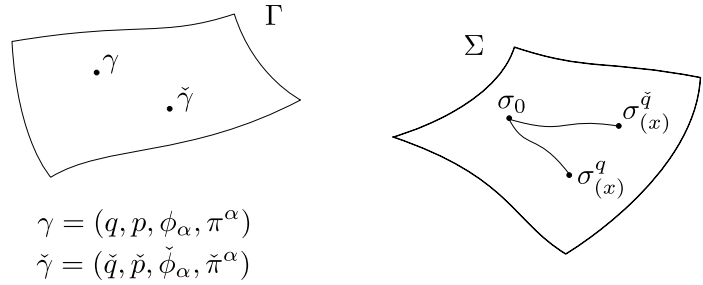}
\caption{The pictures illustrate the building blocks of the separation introduced in equation (\ref{obsdecomp}).\label{obrazek}}
\end{figure}

This separation will play a crucial role in the coming derivations so let us explain it once more on an example. Consider $G(x)$ to be $P^{IJ}(x)$. We want to find a new formula to express its evaluation at a point $\gamma=(q,p,\phi_\alpha,\pi^\alpha)$ (the structures we will be using are depicted in Figure \ref{obrazek}). For the separation (\ref{obsdecomp}) to be possible we need another point in the phase space (the one around which we will variate later), namely $\check\gamma=(\check q,\check p,\check \phi_\alpha,\check \pi^\alpha)$. The first term on the right hand side of (\ref{obsdecomp}) is constructed in the following way: \emph{(i)} Take the field $p$ defining $\gamma$. \emph{(ii)} Express it in coordinates $(\check x^I)$ adapted to $\check q$ defining $\check \gamma$. \emph{(iii)} Evaluate the result at the point $\sigma^{\check q}_{(x)}\in\Sigma$, namely such that the coordinates $\check x^I$ adapted to $\check q$ assign to it the three labels $(x)$. Note, that the only place in which we invoked the phase space point $\gamma$ in the construction of this term was when we chose the field later to be expressed and evaluated. The second term is just the difference between the value of the observable and the first term. In short we will write
\begin{equation}
P^{IJ}(x)\ =\  p^{IJ}(x) + D_{P^{IJ}(x)},
\end{equation}
where, as was already mentioned, the indeces on the two sides of the equality correspond to different sets of coordinates ($(x^I)$ and $(\check x^I)$).

It follows directly from the definition, that     
\begin{equation}\label{Bzeroq}
\left.D_{G(x)}\right|_{(\check q,p,\phi_\alpha,\pi^\alpha)}\ =\ 0.
\end{equation}
This continues to be true for every metric $\widetilde q$ such that the adapted coordinates $(\widetilde x^I)$ coincide with the fixed coordinates $(\check x^I)$ in a neighborhood of the geodesic line connecting points $\sigma_0$ and $\sigma^{\check q}_{(x)}$.
This is true whenever
\begin{equation}\label{q_1Ri}
\widetilde q_{rr}\ =\ 1,\qquad\widetilde q_{rA}\ =\ 0
\end{equation}  
in a neighborhood of  the geodesic line and in the coordinates $(\check y^a)$. For every such metric $\widetilde q$, still
\begin{equation}\label{Bzeroq_1}
\left.D_{G(x)}\right|_{(\widetilde q,p,\phi_\alpha,\pi^\alpha)}\ =\ 0. 
\end{equation}

This observation implies vanishing of a large family of functional derivatives of the functions $D_{G(x)}$. We will discuss that below. On the other hand, the functional derivatives of the fields on the right hand sides of (\ref{obsdecomp}), namely
\begin{equation}
\left.\frac{d}{d\epsilon}\right|_{\epsilon=0}\left.g(x)\right|_{(\check q + \epsilon\delta q, \check p + \epsilon\delta p, \check \phi_\alpha+\epsilon\delta \phi_\alpha, \check \pi^\alpha+\epsilon\delta \pi^\alpha)}
\end{equation}
are straightforward for each of the fields, since they only depend on the varied canonical data through the direct dependance mentioned earlier.

\subsection{Vanishing functional derivatives of $D_{G(x)}$}

In this section we will explicitely discuss the observables defined with the use of the Cartesian adapted coordinates $G(x)$, but all the arguments and results obtained below apply also in the case of the observables $F(r,\theta)$.

\vspace{0.5cm}

Given $G(x)$ and a point $(\check q,\check p,\check \phi,\check \pi)$, we will focus below on the variations of the corresponding function $D_{G(x)}$. The functional derivatives of $D$ are defined by 
\begin{multline}\label{deltaB}
\left.\frac{d}{d \epsilon}\right|_{\epsilon=0}  \left.D_{G(x)}\right|_{(\check q+\epsilon\delta q, \check p+\epsilon\delta p,\check \phi + \epsilon\delta \phi,\check \pi + \epsilon\delta \pi)}\ =\\
=\ \int d^3\sigma \frac{\delta D_{G(x)}}{\delta q_{ij}(\sigma)} \delta q_{ij}(\sigma) + \frac{\delta D_{G(x)}}{\delta p^{ij}(\sigma)} \delta p_{ij}(\sigma) + \frac{\delta D_{G(x)}}{\delta \phi_\alpha(\sigma)} \delta \phi_\alpha(\sigma) + \frac{\delta D_{G(x)}}{\delta \pi^\alpha(\sigma)} \delta \pi^\alpha(\sigma)
\end{multline}
as the map
\begin{equation}\label{dzialanieD}
(\delta q,\delta p,\delta\phi_\alpha,\delta\pi^\alpha)\ \mapsto\ \left.\frac{d}{d \epsilon} \right|_{\epsilon=0} \left.D_{G(x)}\right|_{(\check q+\epsilon\delta q, \check p+\epsilon\delta p,\check \phi+\epsilon\delta \phi_\alpha,\check \pi^\alpha+\epsilon\delta \pi^\alpha)}.
\end{equation}
It follows from (\ref{q_1Ri},\ref{Bzeroq_1}) and the accompanying discussion, that    
\begin{equation}
\frac{d}{d \epsilon} \left.D_{G(x)}\right|_{(\check q+\epsilon\delta q, \check p+\epsilon\delta p,\check \phi+\epsilon\delta \phi_\alpha,\check \pi^\alpha+\epsilon\delta \pi^\alpha)}\ =\ 0
\end{equation}
for all the variations $\delta q, \ldots, \delta \pi^\alpha$ of the fields, such that the condition (\ref{q_1Ri}) is satisfied by the metric
\begin{equation}
\widetilde q\ =\ \check q+\epsilon\delta q
\end{equation}
in the neighborhood of the geodesic connecting $\sigma_0$ with $\sigma^{\check q}_{(x)}$ for every $\epsilon$, that is in the spherical coordinates $(\check y^a)$ adapted to $\check q$, 
\begin{equation} \label{deltaqRi}
\delta q_{rr}\ =\ \delta q_{rA}\ =\ 0, 
\end{equation} 
in the neighborhood of that geodesic. In terms of the functional derivatives on the right hand side of 
(\ref{deltaB}), this observation can be expressed by the following properties:
\begin{itemize}

\item All the non-metric derivatives vanish, namely
\begin{equation}
\frac{\delta D_{G(x)}}{\delta p^{ij}(\sigma)} \ =\ \frac{\delta D_{G(x)}}{\delta \phi_\alpha(\sigma)}\ =\ 
\frac{\delta D_{G(x)}}{\delta \pi^\alpha(\sigma)}\ =\ 0.
\end{equation}

\item The support of the functional derivative
\begin{equation}
\frac{\delta D_{G(x)}}{\delta q_{ij}(\sigma)}
\end{equation}
at $(\check q,p,\phi_\alpha,\pi^\alpha) \in \Gamma$ is contained in the geodesic line connecting $\sigma_0$ with $\sigma^{\check q}_{(x)}$.

\item For every $\delta q$ such that (\ref{deltaqRi}) hold in a neighborhood of the geodesic line connecting $\sigma_0$ with $\sigma^{\check q}_{(x)}$
\begin{equation}\label{condondistr}
\int d^3\sigma \frac{\delta D_{G(x)}}{\delta q_{ij}(\sigma)} \delta q_{ij}(\sigma)\ =\ 0.
\end{equation}
\end{itemize}

Notice, that it follows from the above properties that the $D$ functions Poisson commute with each other
\begin{equation}\label{{BB'}}
\{D_{G(x)},\ D_{G'(x')}\}\ =\ 0.
\end{equation}

\subsection{Diff$_{\text{obs}}$-invariance condition}

Like in the previous sections, we present the reasoning for observables $G(x)$, but all the arguments and results apply also to the observables $F(r,\theta)$.

\vspace{0.5cm}

Another set of identities satisfied by the functional derivatives of the function $D_{G(x)}$ from (\ref{obsdecomp}) follows from the Diff$_{\text{obs}}$-invariance of the observables $G(x)$. In terms of the phase space structure, the observables Poisson commute with the vector constraints
\begin{equation}\label{warunekniezmienniczosci}
\{G(x),\ C(\vec{N})\}\ =\ 0
\end{equation}
for every vector field $\vec{N}$ on $\Sigma$ generating elements of Diff$_{\text{obs}}$, i.e., satisfying conditions (\ref{diffgen}). 

We apply now the decomposition (\ref{obsdecomp}). Because it is $\check q$-dependent, let us spell out the action of diffeomorphisms $\psi\in\text{Diff}_{\text{obs}}$, 
\begin{equation} 
\left.g(x)\right|_{(\psi^*q,\psi^*p,\psi^*\phi_\alpha,\psi^*\pi^\alpha)} + \left.D_{G(x)}\right|_{(\psi^*q,\psi^*p,\psi^*\phi_\alpha,\psi^*\pi^\alpha)}\ =\ 
\left.g(x)\right|_{(q,p,\phi_\alpha,\pi^\alpha)} + \left.D_{G(x)}\right|_{(q,p,\phi_\alpha,\pi^\alpha)},
\end{equation} 
notice that $\psi$ is applied only to $q$ in this formula and it does not affect the dependance of $g$ on $\check q$. Now, we pass from the condition (\ref{warunekniezmienniczosci}) to its differential version at the point $(\check q,\check p,\check \phi_\alpha,\check\pi^\alpha)$ and $\psi$ being the flow of a vector field $\vec{N}$. The condition takes the form of the following equation
\begin{equation}\label{deltaBdeltaq}
\int d^3\sigma \frac{\delta D_{G(x)}}{\delta q_{ij}(\sigma)}{\cal L}_{\vec{N}}{\check q}_{ij}(\sigma)\ =\ 
-\{g(x),\ C(\vec{N})\},
\end{equation}
where the functional derivative is taken at $(\check q,\check p,\check \phi_\alpha,\check\pi^\alpha)$. The Poisson bracket 
on the right hand side is just
\begin{equation} \label{LNF}
\left.\{g(x),\ C(\vec{N})\}\right|_{(\check q,\check p,\check \phi_\alpha,\check\pi^\alpha)}\ =\ 
\left.\frac{d}{d\epsilon}\right|_{\epsilon=0}\left.g(x)\right|_{(\check q+\epsilon {\cal L}_{\vec{N}}\check q, \check p+\epsilon {\cal L}_{\vec{N}}\check p,\check \phi_\alpha+\epsilon {\cal L}_{\vec{N}}\check \phi_\alpha,\check \pi^\alpha+\epsilon {\cal L}_{\vec{N}}\check \pi^\alpha)}.                          
\end{equation}     
For example, for $G(x)=Q_{IJ}(x)$ 
\begin{equation}
\left.\{q_{IJ}(x),\ C(\vec{N})\}\right|_{(\check q,\check p,\check \phi_\alpha,\check\pi^\alpha)}\ =\  {\cal L}_{\vec{N}}{\check q}_{IJ}(x),
\end{equation}
where $I,J$ correspond to coordinates $(\check x^I)$ adapted to $\check q$ and hence
\begin{equation}
\int d^3\sigma \frac{\delta D_{Q_{IJ}(x)}}{\delta q_{ij}(\sigma)}{\cal L}_{\vec{N}}{\check q}_{ij}(\sigma)\ =\ - {\cal L}_{\vec{N}}{\check q}_{IJ}(x).
\end{equation}

\subsection{Decomposition of $\delta q$}

Like in the previous sections, we present the reasoning for observables $G(x)$, but all the arguments and results can be formulated also for the observables $F(r,\theta)$.

\vspace{0.5cm}

Equation (\ref{deltaBdeltaq}) determines the action of the functional  
\begin{equation}\label{actionoffunctional}
\delta q\ \mapsto\  \int d^3\sigma \frac{\delta D_{G(x)}}{\delta q_{ij}(\sigma)}\delta q_{ij}(\sigma)
\end{equation}
on any $\delta q$ that has the form 
\begin{equation}
\delta q = {\cal L}_{\vec{N}}\check q
\end{equation}
with some vector field $\vec{N}$, a generator of the elements of Diff$_{\text{obs}}$. At first sight, that information may seem insufficient to determine the whole distribution (\ref{dzialanieD}). It turns out, however, that every test variation $\delta q$ can be written as
\begin{equation}\label{deltaqdecomp} 
\delta q\ =\ {\cal L}_{\vec{N}}\check q + \delta \widetilde{q} ,     
\end{equation} 
with some vector field $\vec{N}$ as above, and the part $\delta \widetilde{q}$ supported
in the domain of the coordinates $(\check y^a)$ adapted to $\check q$ which satisfies
\begin{equation}
\delta\widetilde{q}_{rr}\ =\ 0\ =\ \delta\widetilde{q}_{rA}.
\end{equation}
But then, a metric $\widetilde q=\check q + \epsilon\delta \widetilde{q}$ satisfies (\ref{q_1Ri}) and in the consequence
\begin{equation}\label{varBtildeq} 
\int d^3\sigma \frac{\delta D_{G(x)}}{\delta q_{ij}(\sigma)}\delta \widetilde{q}_{ij}(\sigma)\ =\  0 
\end{equation}
owing to (\ref{Bzeroq_1}). Therefore, the action of the distribution $\frac{\delta D_{G(x)}}{\delta q}$
on a general test variation $\delta q$ is
\begin{equation}
\int d^3\sigma \frac{\delta D_{G(x)}}{\delta q_{ij}(\sigma)}\delta q_{ij}(\sigma)\ =\  - {\cal L}_{\vec{N}}g(x),
\end{equation}         
where the vector field $\vec{N}$ is determined by the test variation $\delta q$. It can be integrated directly from (\ref{deltaqdecomp}), however, one can also determine it using a general argument and we will do it first in the general way.

Given $\delta q$, for every $\epsilon$ (sufficiently small) consider the metric $\check q + \epsilon \delta q$ and the corresponding adapted coordinates $(x^I_\epsilon)$. Define a diffeomorphism $\psi_\epsilon$ sending the point to which coordinates adapted to $\check q$ give labels $(x)$ to a point to which the coordinates adapted to $\check q + \epsilon \delta q$ assign the same labels, namely
\begin{equation}\label{fepsilon}
\psi_\epsilon : \sigma^{\check q}_{(x)} \mapsto \sigma^{\check q + \epsilon \delta q}_{(x)}
\end{equation} 
For sufficiently small $\epsilon$, the diffeomorphism is well-defined in a neighborhood of the geodesic segment connecting $\sigma_0$ and $\sigma^{\check q}_{(x)}$. The adapted coordinates of the metric tensor $\psi_\epsilon^*(\check q + \epsilon \delta q)$ 
are $(\psi_\epsilon^* x^I_\epsilon)$ and they coincide with the coordinates $(\check x^I)$, in the mentioned neighborhood. Therefore, according to our classification of the vanishing functional derivatives of $D_{G(x)}$ provided above, 
\begin{equation}
\delta \widetilde{q}\ :=\ \left.\frac{d}{d\epsilon}\right|_{\epsilon=0}\psi_\epsilon^*(\check q + \epsilon\delta q)
\end{equation}
satisfies (\ref{varBtildeq}). On the other hand
\begin{equation}
\delta q\ =\ -\left.\frac{d}{d\epsilon}\right|_{\epsilon=0}\psi_\epsilon^* \check q + \left.\frac{d}{d\epsilon}\right|_{\epsilon=0}\psi_\epsilon^*(\check q + \epsilon\delta q)\ =\ {\cal L}_{\vec{N}}\check q + \delta \widetilde{q}
\end{equation}
with
\begin{equation}\label{N}
\vec{N}\ =\ -\left.\frac{d}{d\epsilon}\right|_{\epsilon=0}\psi_\epsilon^*.
\end{equation}

The conclusion is, that the functional derivative $\frac{\delta D_{G(x)}}{\delta q}$ is determined by the equation  (\ref{deltaBdeltaq}) as follows
\begin{equation}\label{conclofaction}
\delta q\ \mapsto \int d^3\sigma\frac{\delta D_{G(x)}}{\delta q_{ij}(\sigma)}\delta q_{ij}(\sigma)\ =\  -{\cal L}_{\vec{N}}g(x),\qquad \text{with }\vec{N}\ =\ -\left.\frac{d}{d\epsilon}\right|_{\epsilon=0}\psi_\epsilon^*
\end{equation}
where $\psi_\epsilon$ is defined by (\ref{fepsilon}) and the right hand side of the first equality is the right hand side of (\ref{LNF}). 

Also, we can conclude, that the equation 
\begin{equation}
\int d^3\sigma T^{ij}(\sigma){\cal L}_{\vec{N}}{\check q}_{ij}(\sigma)\ =\ -\{g(x),\ C(\vec{N})\}
\end{equation}
for an unknown distribution $T^{ij}(\sigma)$ (satisfying (\ref{condondistr})) which is satisfied for every generator $\vec{N}$ of Diff$_{\text{obs}}$ has the unique solution 
\begin{equation}
T^{ij}(\sigma)\ =\ \frac{\delta D_{G(x)}}{\delta q_{ij}(\sigma)}.
\end{equation}

\subsection{Determining $\vec{N}$}

Since in this section we will only be using coordinates adapted to the metric $\check q$ we will simplify the notation by replacing the point at which tensor fields are evaluated, e.g. $\sigma^{\check q}_{(r,\theta)}$, by the values the corresponding coordinates assume, e.g. $(r,\theta)$. So given a tensor field $T$ by $T(r,\theta)$ we will mean the evaluation of the tensor at the point $\sigma^{\check q}_{(r,\theta)}$.

Given a variation $\delta q$ of the metric $\check q$, the vector field $\vec{N}$ appearing in (\ref{deltaqdecomp}), whose existence we have shown above, can be integrated directly from (\ref{deltaqdecomp}). Written in the coordinates $(\check y^a)$, the equations read
\begin{subequations}\label{NR0}
\begin{align}
2N_{r;r}&\ =\ \delta q_{rr},\\
N_{A;r} +N_{r;A}&\ =\ \delta q_{rA}.  
\end{align}
\end{subequations}
Using properties of the spherical adapted coordinates the equations can be cast into the form
\begin{subequations}\label{NR}
\begin{align}
N^{r}_{,r}&\ =\ \frac{1}{2}\delta q_{rr},\label{NRa}\\
N^A_{,r}&\ =\ \check q^{AB}\left(-N^{r}_{,B} + \delta q_{Br}\right).
\end{align}
\end{subequations}
Integrating those equations over $r$ we obtain
\begin{subequations}
\begin{align}
N^{r}(r,\theta)\ &=\ \underset{r\rightarrow 0}{\lim}N^r(r,\theta) + \frac{1}{2}\int_0^r dr'\ \delta q_{rr}(r',\theta),\\
N^A(r,\theta)\ &=\ \underset{r\rightarrow 0}{\lim}N^A(r,\theta) + \int_0^r dr'\ \check q^{BA}(r',\theta) \left(\delta q_{rB}(r',\theta) - \frac{1}{2}\partial_A \left(\int_0^{r'}d r'' \delta q_{rr}(r'',\theta) \right) \right),
\end{align}
\end{subequations}
where we have implemented the simplification of the notation introduced in the first sentence of this section. The second integrand contains the factor $\check q^{AB}$ which behaves as $\frac{1}{r'^2}$ as $r'\rightarrow 0$, however as we explain below the full integrand is finite in $r'=0$. 

Initial conditions at $r=0$  (that is at $\sigma^{\check q}_{(0,\theta)}=\sigma_0$) follow from the conditions on the components  $N^I(0)$ and $N^I_{,J}(0)$ spelled out in (\ref{diffgen}). Analyzing the limit $\underset{r\rightarrow 0}{\lim} \vec{N}(r,\theta)$ we see that the condition on the component $N^r$ reads 
\begin{equation}
\underset{r\rightarrow 0}{\lim}N^r(r,\theta)\ =\ 0.
\end{equation}
The limit of the component $N^A$ is more tricky. To express it we will use the Cartesian coordinates adapted to $\check q$. Although in general a smooth vector field would have divergent limit of the component $N^A$, due to the fact that we require the field to vanish at the origin, the term $\underset{r\rightarrow 0}{\lim}N^A(r,\theta)$ is finite. Using the coordinates regular at the origin we can find the value assumed by $N^A$ in the limit to be
\begin{equation}
\underset{r\rightarrow 0}{\lim}N^A(r,\theta)\ =\ \left(\underset{r\rightarrow 0}{\lim}r\frac{\partial \check y^A}{\partial \check x^I}\right) h^{IJ} n^K \partial_K N_J(0),
\end{equation}
where
\begin{equation}
h^{IJ}\ =\ \delta^{IJ} - n^I n^J
\end{equation}
and all the unit vectors $n^I$ are the ones introduced in (\ref{xtorn}) and they are functions of the angles. Therefore, the limit is a function of the angles.
Introducing a useful notation for a lower triangular matrix $\overline{T}_{IJ}$ which is built from the elements of $T_{IJ}$ in the following way
\begin{equation}
 \begin{bmatrix}
  \overline{T}_{11} & \overline{T}_{12} & \overline{T}_{13}\\ 
  \overline{T}_{21} & \overline{T}_{22} & \overline{T}_{23}\\ 
  \overline{T}_{31} & \overline{T}_{32} & \overline{T}_{33}\\ 
 \end{bmatrix}  
 =
  \begin{bmatrix}
  T_{11} & 0 & 0\\ 
  2T_{21} & T_{22} & 0\\ 
  2T_{31} & 2T_{32} & T_{33}\\ 
 \end{bmatrix}.
\end{equation}
one can (due to (\ref{diffgen}) and (\ref{deltaqdecomp}) expressed in Cartesian coordinates) express the derivatives of $N$ at zero as
\begin{equation}
\partial_I N_J (0)\ =\ \frac{1}{2}\overline{\delta {q}}_{IJ}(0).
\end{equation}
Finally, the resulting $\vec{N}$ (including the contribution from the above limit of $N^A$) is found to be
\begin{multline}\label{resultingN}
 \vec{N}(r,\theta)\ =\ \left[ \frac{1}{2} \overline{\delta q}_{KJ}(0) h^{JL}rn^K \right] \partial_L
+\frac{1}{2}\left[\int_0^r dr'\ \delta q_{rr}(r',\theta)\right] \partial_{r}+\\
+\left[ \int_0^r dr'\ \check q^{BA}(r',\theta) \left(\delta q_{rA}(r',\theta) - \frac{1}{2}\partial_A \left(\int_0^{r'}dr''\ \delta q_{rr}(r'',\theta) \right) \right) \right] \partial_B.
\end{multline}
We would like to note here that although $\check q^{BA}(r',\theta) = O(\frac{1}{r'^2})$ for small $r'$, 
the integral over $r'$ in the last term of the above result is well-defined because, from the identity
\begin{equation}
r\left.\left(\frac{1}{r'}\delta q_{rA}(r',\theta)\right)\right|_{r'\rightarrow 0} - \frac{1}{2}\partial_A \int^r_0 dr'\ \delta q_{rr}(0,\theta)\ =\ rn^I n^J_{,A}\delta q_{IJ}(0) - \frac{1}{2}\partial_A(r n^I n^J \delta q_{IJ}(0))\ =\ 0,
\end{equation}
where $\delta q_{rr}(0,\theta) =\underset{r \rightarrow 0}{\lim} \delta q_{rr}(r,\theta)$, if follows that
\begin{multline}
\delta q_{rA}(r',\theta) - \frac{1}{2}\partial_A \left(\int_0^{r'}dr''\ \delta q_{rr}(r'',\theta) \right)\ = \\
=\ \delta q_{rA}(r',\theta) - r'\left.\left(\frac{1}{r''}\delta q_{rA}(r'',\theta)\right)\right|_{r''\rightarrow 0} - \frac{1}{2}\partial_A \left(\int_0^{r'}dr''\ \delta q_{rr}(r'',\theta) - \delta q_{rr}(0,\theta) \right)\ =\\
=\ O(r'^2).
\end{multline}

Notice also, that in the formula for $\vec{N}$
\begin{equation}
\check q^{BA}(r',\theta)\ =\ Q^{AB}(r',\theta),
\end{equation}
since we are working in a phase space point given by $\check q$ in this section.

The fact that the resulting field (\ref{resultingN}) indeed fulfills the second one of the conditions (\ref{diffgen}) is not that trivial. However, one can confirm it performing an explicit calculation, in which the contributions from the second and third terms cancel most of the contributions from the first term leaving only the expected, lower-triangular, derivative of the field at zero.

\subsection{The result and its meaning \label{theresult}}

Sumarising the results presented in previous sections, for every observable $G(x)$ the distribution $\frac{\delta D_{G(x)}}{\delta q}$ is defined by its action on an arbitrary test variation $\delta q$ as
\begin{equation}
\int d^3\sigma \frac{\delta D_{G(x)}}{\delta q_{ij}(\sigma)}\delta q_{ij}(\sigma)\ =\  - \left.{\cal L}_{\vec{N}}g(\sigma)\right|_{\sigma =\sigma^{\check q}_{(x)}}
\end{equation}         
where the right hand side is just the Lie derivative of the field  $g\in\{q_{IJ},p^{IJ},\phi_\alpha,\pi^\alpha\}$ corresponding to the observable  $G\in\{Q_{IJ},P^{IJ},\Phi_\alpha,\Pi^\alpha\}$ and the indices correspond to the coordinates $(\check x^I)$ adapted to the metric tensor $\check q$ at which the functional derivative of $D_{G(x)}$ is considered. The Lie derivative is then evaluated at the point $\sigma^{\check q}_{(x)}$ which is labelled with the three numbers $x$ by the coordinates $(\check x^I)$.
The vector field $\vec{N}$ is obtained from $\delta q$ according to (\ref{resultingN}).      

Importantly,  the vector field $\vec{N}$ is given explicitly in terms of the coordinates (spherical or Cartesian) adapted to the metric $\check q$ at which the functional derivative $\frac{\delta D_{G(x)}}{\delta q}$ is considered. Therefore, its components can be easily written directly in terms of the values the observables $F(r,\theta)$ evaluated at $q=\check q$, specifically by $\left.Q^{AB}(r,\theta)\right|_{(\check q,p,\phi_\alpha,\pi^\alpha)}$.

Moreover, the derivatives in $\frac{\partial}{\partial y^a}$ and $\frac{\partial}{\partial x^I}$ pass to the derivatives 
of the values of the observables $G(x)$ and $F(r,\theta)$, with respect to the labels $x$ and $r,\theta$. For example, 
\begin{equation}
\left.{\cal L}_{\vec{N}}\phi(\sigma)\right|_{\sigma=\sigma^{\check q}_{(r,\theta)}}\ =\ 
N^A\partial_A\phi(\sigma^{\check q}_{(r,\theta)}) +  N^r\partial_r \phi(\sigma^{\check q}_{(r,\theta)})\ =\ 
\left.N^A\partial_A\Phi(r,\theta)\right|_{q=\check q} + \left.N^r\partial_r \Phi(r,\theta)\right|_{q=\check q}.
\end{equation}
This observation is important for expressing the functional derivatives of the observables themselves by the observables.
To this end it is convenient to consider the space $\mathbb{R}^3$ of the labels $x$ (recall that by $x$ we have collectively denoted the three values the Cartesian adapted coordinates $(x^I)$ assume). In this space we may also use the labelling induced by the spherical adapted coordinates.
For every point in the phase space $(q,p,\phi_\alpha,\pi^\alpha)$, a $q$-dependent neighborhood of $(0,0,0)$ in our label space is endowed with: the metric tensor $Q(x)=Q_{IJ}(x)dx^Idx^J$, the tensor density $x\mapsto P^{IJ}(x)$, the scalar field $x\mapsto\Phi_\alpha(x)$, and the density $x\mapsto \Pi^\alpha(x)$.
Moreover, given a test tensor field $\delta q$ defined in a neighborhood of $\sigma_0\in\Sigma$, we write it in the adapted Cartesian coordinates as
\begin{equation}
\delta q_{IJ}(\sigma^{q}_{(x)})\ =\ w_{IJ}(x)
\end{equation}
and following (\ref{resultingN}) define in the neighborhood of $(0,0,0)$ a vector field $\vec\Lambda$
\begin{multline}\label{NinPois}
\vec\Lambda(x) \ =\ 
 \left[ \frac{1}{2} \overline{w}_{KJ}(0) h^{JL}rn^K\right] \partial_L
+\frac{1}{2}\left[\int_0^r dr'\ w_{rr}(r',\theta)\right] \partial_r +\\
+\left[ \int_0^r dr'\ Q^{BA}(r',\theta) \left(w_{rA}(r',\theta) - \frac{1}{2}\partial_A \left(\int_0^{r'}dr''\ w_{rr}(r'',\theta) \right) \right) \right] \partial_B,
\end{multline}
where $h^{JL}\ =\ \delta^{JL} - n^J n^L$, the $n^I$ are functions of $\theta$ and all $r,\theta$ are determined by $x$ with the use of the (inverse of) the relation (\ref{relationcartspher}).
The vector field $\vec\Lambda$ is continuous and differentiable at $x=(0,0,0)$.   

Given an observable $G(x)$, at a point $(q,p,\phi_\alpha,\pi^\alpha)\in \Gamma$, the functional derivatives are given by
\begin{subequations}\label{thevariation}
\begin{align}
\int d^3x'\ \frac{\delta \Phi_\alpha(x)}{\delta q_{KL}(x')}w_{KL}(x')\ &=\ -\Lambda^K(x)\frac{\partial}{\partial x^K}\Phi_\alpha(x),\\
\int d^3x'\ \frac{\delta \Pi^\alpha(x)}{\delta q_{KL}(x')}w_{KL}(x')\ &=\ -\frac{\partial}{\partial x^K}\left( \Lambda^K(x)\Pi^\alpha(x)\right),\\
\int d^3x'\ \frac{\delta Q_{IJ}(x)}{\delta q_{KL}(x')}w_{KL}(x')\ &=\ w_{IJ}(x) - {\cal L}_{\vec \Lambda} Q_{IJ}(x)\\
\int d^3x'\ \frac{\delta P^{IJ}(x)}{\delta q_{KL}(x')}w_{KL}(x')\ &=\ -{\cal L}_{\vec\Lambda} P^{IJ}(x). 
\end{align}
\end{subequations}

\section{Poisson brackets of the observables}\label{poissonbracketofobs}

\subsection{Poisson brackets of two observables}

A natural question to ask is that about the Poisson brackets of the observables. As one could see already in the formula (\ref{resultingN}), the fixed point $\sigma_0$ (the observer) plays a nontrivial and non-negligible role in the results. This is the reason why, for the sake of precision, it is better to use the observables $G(x)$ defined well at $x=(0,0,0)$ (corresponding to $\sigma=\sigma_0$).
Recall that the transformation (\ref{cyltocart}) between the observables $F(r,\theta)$ and the observables $G(x)$ depends on the (relation of the) labels $x$ and $r,\theta$ while it is independent of the point in the phase space $(q,p,\phi_\alpha,\pi^\alpha)\in\Gamma$, therefore it commutes with the Poisson brackets, for example
\begin{equation}
\{Q_{AB}(r,\theta),\ \cdot\ \}\ =\  
\frac{\partial x^J}{\partial y^A} \frac{\partial x^K}{\partial y^B}\{ Q_{JK}(x^I=r n^I),\ \cdot\ \}.
\end{equation}
     
We will calculate now the Poisson brackets $\{ G(x),\ G'(x')\}$ at a point $(\check q,\check p,\check \phi_\alpha,\check \pi^\alpha)\in \Gamma$, assuming both the observables in the Poisson brackets are well-defined. We will use the spherical coordinates $(\check y^a)$ and the Cartesian coordinates $(\check x^I)$ adapted to $\check q$. We will also make use of the decomposition (\ref{obsdecomp}). The calculation proceeds as follows 
\begin{multline}
\{G(x),\ G'(x')\}\ =\ \{g(x) + D_{G(x)},\ g'(x') + D_{G'(x')}\}\ =\\
= \{g(x),\ g'(x')\} + \{D_{G(x)},\ g'(x')\} - \{D_{G'(x')},\ g(x)\}\ =\\
= \{g(x),\ g'(x')\}
+ \int d^3\sigma\ \frac{\delta D_{G(x)}}{\delta q_{ij}(\sigma)}\frac{\delta g'(x')}{\delta p^{ij}(\sigma)}
- \int d^3\sigma\ \frac{\delta D_{G'(x')}}{\delta q_{ij}(\sigma)}\frac{\delta g(x)}{\delta p^{ij}(\sigma)},  
\end{multline}
where in the first equality we have used (\ref{{BB'}}) and the first bracket in the last line is the usual Poisson bracket between the canonical variables (\ref{ccr}). 

The simplest case is when $G(x)$ and $G'(x')$ do not contain $p$, because then the last two terms above identically vanish:   
 \begin{equation}
\{G(x),\ G'(x')\}\ =\  \{g(x),\ g'(x')\}\qquad\text{for }G,G'\in\{Q_{IJ},\Phi_\alpha,\Pi^\alpha\},   
\end{equation}
that is
\begin{subequations}\label{PBcan}
\begin{align}
\{ \Phi_\alpha(x),\ \Pi^{\alpha'}(x')\}\ &=\ \delta_\alpha^{\alpha'}\delta(x - x'),\\
\{\Phi_\alpha(x),\ \Phi_{\alpha'}(x')\}\ &=\ \{\Pi^\alpha(x),\ \Pi^{\alpha'}(x')\}\ =\ \{Q_{IJ}(x),\ Q_{KL}(x')\}\ =\ 0,\\
\{Q_{IJ}(x),\ \Phi_\alpha(x')\}\ &=\ \{Q_{IJ}(x),\ \Pi^\alpha(x')\}\ =\ 0.
\end{align}
\end{subequations}
Those identities pass to the observables $F(r,\theta)$:
\begin{subequations}\label{PBcanrt}
\begin{align}
\{ \Phi_\alpha(r,\theta),\ \Pi^{\alpha'}(r',\theta')\}\ &=\ \delta_\alpha^{\alpha'}\delta(r - r')\delta(\theta - \theta'),\\
\{\Phi_\alpha(r,\theta),\ \Phi_{\alpha'}(r',\theta')\}\ &=\ \{\Pi^\alpha(r,\theta),\ \Pi^{\alpha'}(r',\theta')\}\ =\ \{Q_{AB}(r,\theta),\ Q_{CD}(r',\theta')\}\ =\ 0,\\
\{Q_{AB}(r,\theta),\ \Phi_\alpha(r',\theta')\}\ &=\ \{Q_{AB}(r,\theta),\ \Pi^\alpha(r',\theta')\}\ =\ 0.
\end{align}
\end{subequations}
The expressions for the Poisson brackets in (\ref{PBcan}) and (\ref{PBcanrt}) are distributions that can be integrated over the labels $x$ and $x'$, and respectively, $r,\theta$ and $r',\theta'$ with arbitrary smearing functions.

Next, consider the bracket $\{G(x),\ \int d^3x'\ w_{IJ}(x')P^{IJ}(x') \} $ for  $G\in \{Q_{IJ},\Phi_\alpha,\Pi^\alpha\}$. It amounts to
\begin{multline}
\{G(x),\ \int d^3x'\ w_{IJ}(x')P^{IJ}(x')\}\ =\\
=\ \int d^3x'\ w_{IJ}(x')\{g(x),\ p^{IJ}(x')\} + \int d^3x'\ \frac{\delta D_{G(x)}}{\delta q_{KL}(x')}w_{KL}(x')\ =\\ 
=\ \int d^3x'\ w_{IJ}(x')\{g(x),p^{IJ}(x')\} - {\cal L}_{\vec\Lambda} G(x).
\end{multline}
The first term of the result contains again the bracket between the canonical variables, while the second term is the Lie derivative acting in the space of labels $x$ on $g\in\{q_{IJ},\phi_\alpha,\pi^\alpha\}$ and acting with respect to the vector field $\vec\Lambda$ defined by (\ref{NinPois}) for the smearing field $w$.
For specific choices of $G$, we have
\begin{subequations}\label{PBPIJ}
\begin{align}
\{\Phi_\alpha(x),\ \int d^3x'\ w_{IJ}(x')P^{IJ}(x')\}\ &=\ - \Lambda^J(x)\frac{\partial}{\partial x^J}\Phi_\alpha(x),\\
\{\Pi^\alpha(x),\ \int d^3x'\ w_{IJ}(x')P^{IJ}(x')\}\ &=\ -\frac{\partial}{\partial x^J}\left( \Lambda^J(x)\Pi^\alpha(x)\right),\\
\{Q_{KL}(x),\ \int d^3x'\ w_{IJ}(x')P^{IJ}(x')\}\ &=\\
=\ w_{KL}(x) - \Lambda^I(x)\frac{\partial}{\partial x^I}&Q_{KL}(x) - \frac{\partial}{\partial x^K}\Lambda^I(x)Q_{IL}(x) - \frac{\partial}{\partial x^L}\Lambda^I(x)Q_{KI}(x).
\end{align}
\end{subequations}
The above expressions are distributions which can be integrated over $x$, with arbitrary smearing functions. To extract from (\ref{PBPIJ}) the Poisson brackets $\{ G(x), P^{AB}(r,\theta)\}$ we write
\begin{equation}
\int d^3x' w_{IJ}(x')\{G(x),\ P^{JK}(x')\}\ :=\ \{G(x),\int d^3x'\ w_{IJ}(x')P^{IJ}(x')\},
\end{equation} 
where $\{G(x),\ P^{JK}(x')\}$ is defined by the right-hand-sides action on an arbitrary test tensor field $w$. Using this object
\begin{equation}
\{ G(x),\ P^{AB}(r,\theta)\}\ =\ \left|\det\left(\frac{\partial x'}{\partial y}\right)\right| \frac{\partial y^A}{\partial x'^I}\frac{\partial y^B}{\partial x'^J}\{G(x),\ P^{IJ}(x')\}.
\end{equation} 

Denote by $T_{IJ}^{KL}$, a constant matrix such that 
\begin{equation}
\overline{w}_{KL}(0)\ =\ T_{KL}^{IJ}w_{IJ}(0). 
\end{equation}
The following Poison brackets follow from (\ref{PBPIJ}), 
\begin{subequations}\label{PBPAB}
\begin{align}
\{ \Phi_\alpha(r,\theta),\ P^{AB}(r',\theta')\}\ =&\ -\frac{1}{2} y'^A_{,I} y'^B_{,J} T^{IJ}_{KL} h^{LM}x^K y^C_{,M} \partial_C\Phi_\alpha(r,\theta)\delta(r'),\\
\{ \Pi^\alpha(r,\theta),\ P^{AB}(r',\theta')\}\ =&\ -\frac{1}{2} y'^A_{,I} y'^B_{,J} T^{IJ}_{KL} \left( h^{LM} x^K y^C_{,M}\Pi^\alpha(r,\theta)\right)_{,C}\delta(r'),\\
\{Q_{CD}(r,\theta),\ P^{AB}(r',\theta')\}\ =&\ \delta^A_{(C} \delta^B_{D)} \delta(r - r')\delta(\theta - \theta')+\\
&-\frac{1}{2} y'^A_{,I}y'^B_{,J} T^{IJ}_{KL} h^{LM} x^K y^E_{,M}Q_{CD,E}(r,\theta)\delta(r')+\\
&-\frac{1}{2} y'^A_{,I}y'^B_{,J} T^{IJ}_{KL} \left( h^{LM} x^K y^E_{,M}\right)_{,C}Q_{ED}(r,\theta)\delta(r')+\\
&-\frac{1}{2} y'^A_{,I}y'^B_{,J} T^{IJ}_{KL} \left( h^{LM} x^K y^E_{,M}\right)_{,D}Q_{CE}(r,\theta)\delta(r').
\end{align}
\end{subequations}
Note that for the sake of brevity we noted fields depending on the $r,\theta$ labels without a prime (e.g. $h^{LM}$), while the ones depending on $r',\theta'$ are denoted with a prime (e.g. $y'^A_{,I}$). The functions $y'^A_{,I}$ depend on $r'$ as $\frac{1}{r'}$, however, the asymptotic behavior as $r'\rightarrow 0$ of the $w_{AB}(r',\theta')$ components of a smooth test tensor field behave as $r'^2$.  

In a similar way one can calculate $\{ F(r,\theta),\ P^{rr}(r',\theta')\}$ and $\{ F(r,\theta),\ P^{rA}(r',\theta') \}$
for $F\in\{\Phi_\alpha,\Pi^\alpha,Q_{AB}\}$. 

The last Poisson bracket is  
\begin{multline}\label{PIJPIJ}
\{\int d^3x\ w_{IJ}(x)P^{IJ}(x),\ \int d^3x'\ w'_{KL}(x')P^{KL}(x')\}\ =\\
=\ \int d^3x\ w_{IJ}(x)\int d^3x'\frac{\delta D_{P^{IJ}(x)}}{\delta q_{KL}(x')}w'_{KL}(x') - \int d^3x'\ w'_{KL}(x') \int d^3x\ \frac{\delta D_{P^{KL}(x')}}{\delta q_{IJ}(x)}w_{IJ}(x)\ =\\
=\ \int d^3x\ \left(w_{IJ}(x){\cal L}_{\vec\Lambda'} - w'_{IJ}(x){\cal L}_{\vec\Lambda}\right)P^{IJ}(x)),
\end{multline} 
where  ${\cal L}$ is the Lie derivative in the space of the labels $x$, the vector field $\vec\Lambda$ is the one deifned by (\ref{NinPois}) while the vector field $\vec\Lambda'$ is obtained by replacing $w$ with $w'$ in (\ref{NinPois}).
As above, the Poisson bracket $\{P^{AB}(r,\theta),\ P^{CD}(r',\theta')\}$ can be obtained from that result and it will be the sum of terms proportional to either $\delta(r)$ or $\delta(r')$.

\subsection{The Poisson bracket of the observables with the vector constraint}\label{actionofdiffeo}

Each observable $F(r,\theta)$ we have defined satisfies
\begin{equation}
\{F(r,\theta),\ C(\vec N)\}\ =\ 0
\end{equation} 
for every vector field $\vec N$ on $\Sigma$ satisfying conditions (\ref{diffgen}). What about a general vector field $\vec M$
defined on the manifold $\Sigma$?
We will answer this question in this section. 

To calculate the Poisson bracket $\{F(r,\theta),\ C(\vec M)\}$ at a point of the phase space $(q,p,\phi,\pi)\in\Gamma$ we will use the  assigned orthonormal frame $(e^0_I)$ in $T_{\sigma_0}\Sigma$, the adapted spherical coordinates $(y^a)$ and the adapted Cartesian coordinates $(x^I)$ defined in a neighborhood of $\sigma_0$. For the observables we will use the decomposition (\ref{obsdecomp}).

\subsubsection{A general consideration}\label{ageneralconsideration}

Using the (\ref{obsdecomp}) decomposition
\begin{equation}\label{1}
\{F(r,\theta),\ C(\vec M)\}\ =\  \{f(r,\theta),\ C(\vec M)\} + \{D_{F(r,\theta)},\ C(\vec M)\}.  
\end{equation}
Given $F$, the first term is known to be the Lie derivative of $f$ with respect to the field $\vec M$. In the second term we will apply our results  concerning the variations of $D_{F(r,\theta)}$ 
\begin{equation}
\{D_{F(r,\theta)},\ C(\vec M)\}\ =\ \int d^3\sigma\ \frac{\delta D_{F(r,\theta)}}{\delta q_{ij}(\sigma)} \frac{\delta C(\vec M)}{\delta p^{ij}(\sigma)}\ =\  \int d^3\sigma\ \frac{\delta D_{F(r,\theta)}}{\delta q_{ij}(\sigma)} 2M_{(i;j)}(\sigma).  
\end{equation}
The last integral has exactly the form (\ref{actionoffunctional}) considered above with
\begin{equation}\label{4}
\delta q_{ij}\ :=\ 2M_{(i;j)}.
\end{equation}
Therefore if $N$ is the vector field defined by (\ref{deltaqdecomp}) for the above $\delta q$, then according
to (\ref{conclofaction})
\begin{equation}\label{2}
\int d^3\sigma\ \frac{\delta D_{F(r,\theta)}}{\delta q_{ij}(\sigma)} 2M_{(i;j)}(\sigma)\ =\ -\{f(r,\theta),\ C(\vec N)\}. 
\end{equation}
Combining (\ref{1}) and (\ref{2}) we conclude that
\begin{equation}\label{3}
\{F(r,\theta),\ C(\vec M)\}\ =\ \{f(r,\theta),\ C(\vec \Delta)\}\ =\ \left.{\cal L}_{\vec\Delta}f(\sigma)\right|_{\sigma=\sigma^q_{(r,\theta)}}, 
\end{equation}
where $\vec\Delta = \vec M -\vec N$. What is left to do is, given a vector field $\vec M$, to find the corresponding vector field $\vec\Delta$.

\subsubsection{Equation determining $\vec\Delta$}

Invoking (\ref{NR0}) the vector field $\vec N$ corresponding to $\delta q$ from (\ref{4}) satisfies equations 
\begin{subequations}
\begin{align}
N_{r;r}\ &=\ M_{r;r},\\
N_{r;A}+N_{A;r}\ &= \ M_{r;A}+M_{A;r}. 
\end{align}
\end{subequations}
hence, the vector field $\vec\Delta$ satisfies the following equations
\begin{subequations}
\begin{align}
\Delta_{r;r}\ &=\ 0,\\
\Delta_{r;A} + \Delta_{A;r}\ &=\ 0.  
\end{align}
\end{subequations}
Those equations can be recast in the form
\begin{subequations}\label{ab}
\begin{align}
\partial_r\Delta^r\ &=\ 0,\label{a}\\
\partial_r\Delta^A\ &= \ -q^{AB}\partial_B\Delta^r.\label{b}
\end{align}
\end{subequations}
The initial conditions the vector field $\vec N$ satisfies at $\sigma_0$ are given by (\ref{diffgen}). In our case (\ref{4}) hold, so the initial conditions read
\begin{subequations}
\begin{align}
N^I(\sigma_0)\ &=\ 0,\\
\partial_{I}N^J(\sigma_0)\ &=\ \begin{cases}M_{I,J}(\sigma_0) \qquad\text{for } I=J\\
                                                       2M_{(I,J)}(\sigma_0) \quad\text{for } I<J\\
                                                       0 \ \ \quad\qquad\qquad\text{for } I>J .
                                                       \end{cases}    
\end{align}
\end{subequations}
They imply the following initial conditions for $\vec\Delta$ at $\sigma_0$
\begin{subequations}\label{DeltaM0}
\begin{align}
\Delta^I(\sigma_0)\ &=\ M^I(\sigma_0),\label{Deltawzerzewarunek}\\
\partial_{I}\Delta^J(\sigma_0)\ &=\ \begin{cases}0 \qquad\qquad\quad\text{for } I=J\\\label{Deltawarunekwzerzepochodna}
                                                       -M_{I,J}(\sigma_0) \ \quad\text{for } I<J\\
                                                       M_{J,I}(\sigma_0)\quad\text{for } I>J .
                                                       \end{cases}    
\end{align}
\end{subequations}

\subsubsection{The case $\vec M(\sigma_0)=0$}

In this case a general solution to the equations (\ref{ab}) is
\begin{subequations}
\begin{align}
\Delta^r\ &=\ 0,\\
\Delta^A(\sigma_{(r,\theta)})\ &=\ \Delta^A(\sigma_{(0,\theta)}),\label{DeltaAwzerze}
\end{align}
\end{subequations}
that is
\begin{equation}
\vec \Delta\ =\ \Delta^A(\sigma_0)\partial_A,
\end{equation}
with a priori arbitrary finite $\Delta^A(\sigma_0)$ being functions of the angles only.

To ensure the initial conditions in $\sigma_0$ we will express $\Delta^A$ by the components $\Delta^I$ 
in the Cartesian adapted coordinates $(x^I)$, and use
\begin{equation}
\frac{\partial y^A}{\partial x^I}\ =\ \frac{1}{r}t^A_I,
\end{equation}
where $t^A_I$ is a function of the angles $\theta$, but is independent of $r$. Now,
\begin{equation}
\Delta^A\ =\ \frac{\partial y^A}{\partial x^I}\Delta^I\ =\ t^A_I\frac{\Delta^I}{r},
\end{equation}
but from (\ref{DeltaAwzerze}) we know that the left hand side does not depend on $r$ so we can introduce a limit in the following way
\begin{equation}
\Delta^A\ =\ \ t^A_I\underset{r\rightarrow0}{\lim}\frac{\Delta^I}{r}
\end{equation}
and since we are considering a case in which $M^I(\sigma_0)=0$, invoking (\ref{Deltawzerzewarunek}) we find
\begin{equation}
\Delta^A\ =\ t^A_I n^J \partial_J\Delta^I(\sigma_0).
\end{equation}
Finally, in a neighborhood of $\sigma_0$,
\begin{multline}\label{resultingDelta}
\vec\Delta\ =\  \Delta^A\partial_A\ =\ \left[n^J  t^A_I \partial_J \Delta^I(\sigma_0)\right] \partial_A\ =\\
=\ \left[x^J \left(\delta_I^K - n_I n^K\right) \partial_J \Delta^I(\sigma_0)\right]\partial_K\ =\ \left[x^J\partial_J\Delta^K(\sigma_0)\right]\partial_K.
\end{multline} 
Notice that this result means that in the case $M^I(\sigma_0)=0$ the field $\vec\Delta$ is fully determined by the three entries appearing above the diagonal of $\partial_I M^J(\sigma_0)$.

\subsubsection{The case $\vec M(\sigma_0) \neq 0$}

A general solution to the equation (\ref{a}) is
\begin{equation}
\Delta^r(\sigma_{(r,\theta)})\ =\ \Delta^r(\sigma_{(0,\theta)})\ =\ \Delta^I(\sigma_0)n_I,
\end{equation}
where, due to (\ref{Deltawzerzewarunek}), $\Delta^I(\sigma_0)$ is a given initial value.

With the above solution, the equation (\ref{b}) reads
\begin{equation}
\partial_r\Delta^A(\sigma_{(r,\theta)})\ = \ -q^{AB}(\sigma_{(r,\theta)})\partial_B n_K(\theta)\Delta^K(\sigma_0).\label{b'}
\end{equation}    
The components $q^{AB}(\sigma_{(r,\theta)})$ have the asymptotics in $r=0$ of the type $\frac{u(\theta)}{r^2}$,
\begin{equation}
q^{AB}(\sigma_{(r,\theta)})\ =\ \frac{\partial y^A}{\partial x^I} \frac{\partial y^B}{\partial x^J}q^{IJ}(\sigma_{(r,\theta)})\ =\ t^A_I(\theta)t^B_J(\theta) \frac{q^{IJ}(\sigma_{(r,\theta)})}{r^2},
\end{equation}
where $q^{IJ}$ are components of the metric tensor in the Cartesian adapted coordinates $(x^I)$. Since the first two leading terms will be relevant for us, note that $q^{IJ}$ has the following expansion
\begin{equation}
q^{IJ}(\sigma_{(r,\theta)})\ =\ \delta^{IJ} + s^{IJ}(r,\theta)r^2,
\end{equation} 
where $s^{IJ}(r,\theta)$ for every fixed $\theta$ is a function of $r$ finite at $0$. Applying this expansion in equation (\ref{b'}) we get
\begin{equation}\label{b''}
\partial_r\Delta^A(\sigma_{(r,\theta)})\ = \ - t^A_I(\theta) h_{KJ}(\theta)\Delta^K(\sigma_0)\left(\frac{\delta^{IJ}}{r^2} + s^{IJ}(r,\theta)\right),
\end{equation} 
where we denoted
\begin{equation}
h_{KJ}(\theta)\ =\ \delta_{KJ} - n_K(\theta)n_J(\theta)
\end{equation}
and explicitly pointed out the dependence on $r$ and $\theta$. 

The general solution to (\ref{b''}) is defined up to $C^A$, a function of $\theta$ constant in $r$,
\begin{equation}
\Delta^A(\sigma_{(r,\theta)})\ =\ t^A_I(\theta) h_{KJ}(\theta)\Delta^K(\sigma_0)\left(\frac{\delta^{IJ}}{r} - \int_0^r dr'\ s^{IJ}(r',\theta)\right) + C^A(\theta).\label{b'''}
\end{equation} 

For a general, continuous and differentiable in $\sigma_0$ vector field $\vec V$, the expansion in $r$ near $\sigma_0$ of its angular components is
\begin{equation}
V^A(\sigma_{(r,\theta)})\ =\ \frac{\partial y^A}{\partial x^I}V^I(\sigma_{(r,\theta)})\ =\ t^A_I(\theta)\left( \frac{V^I(\sigma_0)}{r} + \partial_r V^I(\sigma_0) + O(r)\right).
\end{equation}
The comparison of this expansion with (\ref{b'''}) determines $C^A$ to be
\begin{equation}
C^A(\theta)\ =\ t^A_I(\theta) n^J(\theta) \partial_J\Delta^I(\sigma_0). 
\end{equation} 

The resulting solution in a comprehensive form reads
\begin{equation}
\vec\Delta\ =\ \Delta^I(\sigma_0) \partial_I + x^J\partial_J\Delta^I(\sigma_0)\partial_I + h_{KJ}\Delta^K(\sigma_0) 
\left(r \int_0^r dr'\ s^{IJ}(r',\theta) \right) \partial_I.
\end{equation}

\subsubsection{Interpretation of the action of $\vec M=\vec N + \vec\Delta$}\label{interpretsection}

Given a metric tensor $q$ on $\Sigma$, we have decomposed a general vector field $\vec M$ defined on $\Sigma$ into the sum of $\vec N$ and $\vec\Delta$. The flow of $\vec N$ is contained in the subgroup Diff$_{\text{obs}}$ of the diffeomorphisms which preserve our observables. The vector field $\vec\Delta$, on the other hand, infinitesimally preserves the radial form of the metric $q$. That is, if $\psi_t$ is an element of the flow of $\vec\Delta$ in $\Sigma$, then
\begin{equation}\label{ftq}
\psi_t^*q\ =\ \psi_t^*\left(dr\otimes dr + q_{AB}d\theta^A\otimes d\theta^B\right)\ =\ dr\otimes dr + (q_{AB} + tq'_{AB})d\theta^A\otimes d\theta^B + O(t^2).
\end{equation}
Given $q$, the space ${\cal S}_q$ of all the possible vector fields $\vec V$ whose flows satisfy (\ref{ftq}) is parametrized by $6$ free numbers: 
$3$ components $V^I(\sigma_0)$ and $3$ independent entries of the necessarily antisymmetric $\partial_I V^J(\sigma_0)$. Moreover, ${\cal S}_q$ is a vector space and each $\vec V\in {\cal S}_q$ can be obtained as the vector field $\vec\Delta$ for some choice of $\vec M$. There is a difference from the point of view of the preservation of the radial form between the cases $\Delta^I(\sigma_0)=0$,
and respectively $\Delta^I(\sigma_0)\neq0$. 

In the first case, the vector field $\vec\Delta$ depends on the given metric $q$ only through the adapted coordinates (spherical or Cartesian). It can be described as defined by a single vector field
\begin{equation}
\vec V_T\ =\  V_T^A(\theta)\partial_A
\end{equation}
tangent to the unit sphere in the tangent space $T_{\sigma_0}\Sigma$ pushed forward by the exponent map to all the spheres $r=\text{const}$. While every vector field $\vec V_T$ defines in that way a vector field in $\Sigma$ continuous in $\sigma_0$, the resulting vector field might in general turn out not to be differentiable at $\sigma_0$. It is easy to show, that the necessary and sufficient condition for the differentiability of $\vec V_T$ is that it should be a generator of a rotation of the sphere with respect to the metric $q(\sigma_0)$. This is exactly the meaning of the condition (\ref{Deltawarunekwzerzepochodna}) on $\partial_J\Delta^I(\sigma_0)$ we have derived. Furthermore, in the case $\Delta^I (\sigma_0)=0$, the preservation law (\ref{ftq}) is satisfied for every metric tensor $q'$ which in the coordinates $(y^a)$ adapted to the metric $q$ also takes the radial form 
\begin{equation}\label{q'}
q'\ =\ dr\otimes dr + q'_{AB}d\theta^A\otimes d\theta^B.
\end{equation}
For that reason, the commutator of two vector fields $\vec\Delta,\vec\Delta'\in {\cal S}_q$ vanishing in $\sigma_0$ is also a vector field $\vec\Delta''\in{\cal S}_q$. Therefore, those vector fields can be integrated to a group of local diffeomorphisms $\psi: {\cal U}\mapsto {\cal U}$ acting in a domain ${\cal U}\subset \Sigma$ of a given system of Cartesian coordinates adapted to $q$ such that 
\begin{equation}\label{f1}
x^I(\psi(\sigma))\ =\ R^I_J x^J(\sigma),
\end{equation}
where $R^I_J$ is a fixed matrix of a rotation in $\mathbb{R}^3$. 
 
In the  case $\Delta^I(\sigma_0)\neq0$, on the other hand, the vector field $\vec\Delta$ depends not only on the adapted coordinates of the metric $q$, but also on the components $q_{AB}$. Therefore, the infinitesimal preservation law (\ref{ftq}) holds for the metric $q$ and in general does not hold for another $q'$ which also takes the radial form (\ref{q'}). Those infinitesimal diffeomorphisms can be also related to finite local diffeomorphisms $\psi:{\cal U}\rightarrow {\cal U}$ preserving the radial form of the metric $q$. Each local diffeomorphism of that property, given $q$, can be defined by a point $\sigma'_0$, a frame $(e'^0_I)$ at $\sigma'_0$ (orthonormal frame $e'_I$ is defined as before by Gram-Schmidt orthonormalization of $e'^0_I$ but now with respect to metric $q(\sigma')$). Using the frame $(e'_I)$ we introduce the Cartesian coordinates $(x'^I)$ adapted to $q$ in a neighborhood of $\sigma'_0$. 
For every $\sigma$ in that neighborhood, $\psi(\sigma)$ is the point for which
\begin{equation}\label{f2}
x'^I(\psi(\sigma))\ =\ x^I(\sigma).
\end{equation}
In particular 
\begin{equation}
\psi(\sigma_0)\ =\ \sigma'_0\qquad \text{and}\qquad T\psi(\sigma_0)e_I = e'_I.
\end{equation}
In case $\sigma_0=\sigma'_0$ we can take $e'_I=R^J_I e_J$. Then $R^I_J x'^J(\sigma)\ =\  x^I(\sigma)$ and (\ref{f2}) becomes (\ref{f1}).

\section{Application: dynamics and the constraints} 

Consider a theory defined in our original phase space $\Gamma$ of Section \ref{sectionFandF} (before the observer was introduced) by the vector constraints 
\begin{equation}
C_i(\sigma)\ =\ 0
\end{equation}
considered above, and by a Hamiltonian
\begin{equation}
H\ =\ \int_\Sigma d^3\sigma \left(h(q,p,\phi_\alpha,\pi^\alpha)(\sigma) + N^i(\sigma) C_i(\sigma)\right)
\end{equation}
such that  
\begin{equation}
\{ \int_\Sigma d^3\sigma h(q,p,\phi_\alpha,\pi^\alpha)(\sigma),\ C_i(\sigma)\}\ =\ 0,
\end{equation}
where $h(\sigma)$ depends on the values of the fields $q,p,\phi_\alpha,\pi^\alpha$ and their derivatives at $\sigma$.
An example of such a theory is given by gravity coupled to other fields and deparametrized by one of them. In that deparametrization framework the scalar constraint is solved with respect to the momentum canonically conjugate to a distinguished scalar field, while the scalar field itself is 
eliminated by a suitable gauge choice. The fields $(q,p,\phi_\alpha,\pi^\alpha)$ featuring in the definition of our phase space are in that case the remaining fields.  Specifically, the deparametrising field may be a massless Klein-Gordon field (the Rovelli-Smolin theory \cite{RovelliSmolin}) or the scalar field of Brown-Kucha$\check{\rm r}$ describing an irrotational dust \cite{BrownKuchar, HusainPawlowski, Swiezewski}. In those cases the hamiltonian density is
\begin{subequations}\label{RSandBK}
\begin{multline}
h_{\rm RS}(p,q,\phi_\alpha,\pi^\alpha)\ =\\
=\ \pm \sqrt{\sqrt{\det q}\left(-C^{\rm gr} - C^{\rm matt} \pm \sqrt{\left(C^{\rm gr} + C^{\rm matt}\right)^2 - q^{ij}\left(C_i^{\rm gr} + C_i^{\rm matt}\right)\left(C_j^{\rm gr} + C_j^{\rm matt}\right)}\right)},
\end{multline}
and, respectively,
\begin{equation}
h_{\rm BK}(p,q,\phi_\alpha,\pi^\alpha)\ =\ C^{\rm gr} + C^{\rm matt},
\end{equation}
\end{subequations}     
where $C^{\rm gr}$ is the usual ADM scalar constraint, $C^{\rm gr}_i$ is the usual ADM vector constraint (both depending on $q$ and $p$), whereas $C^{\rm matt}$ and $C^{\rm matt}_i$ are the respective contributions from matter fields (depending also on the fields $\phi_\alpha$ and $\pi^\alpha$).

The group of the gauge transformations consists of the group of transformations of the phase space $\Gamma$ induced by the group of the diffeomorphisms of $\Sigma$ (having a compact support in $\Sigma$). Our observables $G(x)$ and $F(r,\theta)$ are almost invariant with respect to the gauge transformations. They are subject only to the 6 dimensional family of the residual gauge transformations characterized in Section \ref{actionofdiffeo}.      

In this section we will study the dynamics of our observables and use them to solve the vector constraints.

\subsection{Dynamics}\label{dynamo}

The dynamics of our Diff$_{\text{obs}}$-invariant observables is defined by a Poisson bracket with the Hamiltonian $H$. We calculate it at a point $(q,p,\phi_\alpha,\pi^\alpha)\in \Gamma$ using the adapted spherical coordinates $(y^a)$, adapted Cartesian coordinates $(x^I)$ and the decomposition (\ref{obsdecomp})
\begin{equation}
G(x)= g(x) + D_{G(x)},\qquad F(r,\theta)\ =\ f(r,\theta) + D_{F(r,\theta)}.
\end{equation}
That is, 
\begin{equation}\label{GsigmaH}
\{ G(x),\ H \}\ =\ \{g(x),\ H\} + \int d^3\sigma'\ \frac{\delta D_{G(x)}}{\delta q_{ij}(\sigma)}\frac{\delta H}{\delta p^{ij}(\sigma)}\ =\ \{g(x),\ H\} - ({\cal L}_{\vec\Lambda} g)(x), 
\end{equation}
where $\vec\Lambda$ is the vector field defined by (\ref{NinPois}) with 
\begin{equation}
w_{IJ}(\sigma)\ =\ \frac{\delta H}{\delta p_{IJ}(\sigma)}.
\end{equation}
It is important, from the point of view of the framework, to express the Poisson bracket $\{ G(x),\ H \}$ by the observables themselves. Before doing that, notice that it is typically true (including the examples (\ref{RSandBK})) that the first term in (\ref{GsigmaH}) - the Poisson bracket with one of the canonical variables $g(x)$ has the following form
\begin{equation} 
\{g(x),\ H\}\ =\ H_1\left(q_{ij}(\sigma),q_{ij,k}(\sigma),q_{ij,kl}(\sigma),p^{ij}(\sigma),\phi_\alpha(\sigma),\phi_{\alpha,i}(\sigma),\pi^\alpha(\sigma)\right)
\end{equation}
with some function $H_1$ and an arbitrary coordinate system $(z^i)$ on $\Sigma$ (recall that $g(x)$ does not depend on the metric through  the use of adapted coordinates - see the discussion at the begining of Section \ref{yetanotherformula}). Given a point $(q,p,\phi_\alpha,\pi^\alpha)\in\Gamma$, we can use the Cartesian adapted coordinates $(x^I)$ and the corresponding spherical adapted coordinates $(y^a)$,
\begin{equation}
\{g(x),\ H\}\ =\ H_1\left(Q_{IJ}(x),Q_{IJ,K}(x),Q_{IJ,KL}(x),P^{IJ}(x),\Phi_\alpha(x),\Phi_{\alpha,I}(x),\Pi^\alpha(x)\right).
\end{equation}
Now,
\begin{multline}
\{ G(x),\ H \}\ =\\
=\ H_1\left(Q_{IJ}(x),Q_{IJ,K}(x),Q_{IJ,KL}(x),P^{IJ}(x),\Phi_\alpha(x),\Phi_{\alpha,I}(x),\Pi^\alpha(x)\right) - \left({\cal L}_{\vec\Lambda} G\right)(x),   
\end{multline}
where $w_{IJ}$ used in the definition $\vec\Lambda$ is also expressed by the observables $G(x)$ 
\begin{equation}
w_{IJ}(x)\ =\ W_{IJ}\left(Q_{IJ}(x),Q_{IJ,K}(x),Q_{IJ,KL}(x),P^{IJ}(x),\Phi_\alpha(x),\Phi_{\alpha,I}(x),\Pi^\alpha(x)\right).
\end{equation}
At this point, the derivatives featuring in the arguments of the functions $H_1$ and $W$ are considered as derivatives  with respect to the labels $x$, labeling the observables. The conclusion is, that the time evolution $\frac{d}{d t}G(x)$ of any of the observables corresponding to a given $x$, contains terms proportional to:
\begin{itemize}
\item observables $G'(x)$, and their derivatives $G'_{,I}(x)$, $G'_{,IJ}(x)$, ...
\item observables $G'(0)$, and their derivatives $G'_{,I}(0)$, $G'_{,IJ}(0)$ ... 
\item integrals  $\int_0^1 d\tau\ l(G'(\tau x),G'_{,I}(\tau x),G'_{,IJ}(\tau x), ...)$ along the line connecting $x$ with $0$
(which in $\Sigma$ corresponds to the geodesic interval connecting $\sigma$ with $\sigma_0$).
\end{itemize}

Another natural question is whether the Hamiltonian itself can be written as or replaced by a function of the observables. Suppose, in a neighborhood of a given $(q,p,\phi_\alpha,\pi^\alpha)\in\Gamma$, the Cartesian adapted coordinates $(x^I)$ are defined globally on $\Sigma$ (this is true, e.g., for $q$'s sufficiently close to being flat). Then, the Hamiltonian can be expressed by the observables    
\begin{multline}
\int_\Sigma d^3\sigma\ h\left(q_{ij}(\sigma),q_{ij,k}(\sigma),q_{ij,kl}(\sigma), p^{ij}(\sigma),\phi_\alpha(\sigma),\phi_{\alpha,i}(\sigma),\pi^\alpha(\sigma)\right)\ =\\
=\ \int d^3x\ h\left(Q_{IJ}(x),Q_{IJ,K}(x),Q_{IJ,KL}(x),P^{IJ}(x),\Phi_\alpha(x),\Phi_\alpha(x),\Pi^\alpha(x)\right).
\end{multline}
In such a case, the Poisson bracket is automatically given by the Poisson brackets $\{G(x),\ G'(x')\}$,  
\begin{multline}\label{dynobs}
\{G(x),\ H\}\ =\ \int d^3x'\ \left[\frac{\delta H}{\delta Q_{IJ}(x')}\{G(x),\ Q_{IJ}(x')\} + \frac{\delta H}{\delta P^{IJ}(x')}\{G(x),\ 
P^{IJ}(x')\}\right. + \\
+ \left.\frac{\delta H}{\delta \Phi_\alpha(x')}\{G(x),\ \Phi_\alpha(x')\} + \frac{\delta H}{\delta \Pi^\alpha(x')}\{G(x),\ \Pi^\alpha(x')\} \right]. 
\end{multline}
In a general point $(q,p,\phi_\alpha,\pi^\alpha)\in\Gamma$ the observables $G(x)$ are defined only for $x$ from some 
neighborhood of 0 in ${\mathbb R}^3$. The best we can do is to introduce quasi-local hamiltonians,        
\begin{equation}
H_{\cal U}\ =\ \int_{\cal U} d^3x\ h\left(Q_{IJ}(x),Q_{IJ,K}(x),Q_{IJ,KL}(x),P^{IJ}(x),\Phi_\alpha(x),\Phi_{\alpha,I}(x),\Pi^\alpha(x)\right).
\end{equation}
defined manifestly by the observables $G(x)$.  
Then, for every $x'\in \mathbb{R}^3$ such that the line connecting $x'$ and 0 is contained in $\cal U$ and each observable $G(x')$, we have
\begin{equation}
\{G(x'),\ H\}\ =\ \{G(x'),\ H_{\cal U}\}. 
\end{equation}

\subsection{Solutions of the vector constraint} \label{sekcjarozwdiff}

Our definition (\ref{Pirtheta}) of the Diff$_{\text{obs}}$-invariant observables is valid also for the components $P^{rr}$ and $P^{rA}$ of the gravitational field momentum. Observables $P^{rr}(r,\theta)$ and $P^{rA}(r,\theta)$ were included in (\ref{obs}), but as we already mentioned, on the vector constraint subspace $\Gamma_{C}\subset\Gamma$, which is distinguished by the vanishing of the vector constraint, we can express the observables $P^{rr}(r,\theta)$ and $P^{rA}(r,\theta)$ by the remaining observables from (\ref{obs}). We discuss this in more detail now, because the observables $P^{ra}(r,\theta)$ (as functions of (\ref{obs})) are present in the hamiltonians from the previous section.

For every point $(q,p,\phi_\alpha,\pi^\alpha)\in\Gamma_C$ the components of $q$, $p$ and the fields $\phi_\alpha$, $\pi^\alpha$ satisfy the vector constraint at every point $\sigma\in\Sigma$. The constraint reads   
\begin{equation}\label{vce}
C_{i}(\sigma)\ =\ -2\nabla_{j}p_{\phantom{j}i}^{j} + C^{\rm matt}_i(\phi_\alpha,\pi^\alpha)\ =\ 0,
\end{equation}
where $C^{\rm matt}_i$ is the term contributed by the fields $\phi_\alpha$ and their momenta $\pi^\alpha$. Let us use the spherical coordinates $(y^a)$ adapted to $q$. We will consider equation (\ref{vce}), as an equation for some of the momenta observables $P^{ab}(r,\theta)$, given by the remaining momenta, the metric observables $Q_{ab}(r,\theta)$ and matter fields observables $\Phi_\alpha(r,\theta)$ and $\Pi^\alpha(r,\theta)$. Let us first discuss its properties using just the fields $q,\ldots,\pi^\alpha$.
  
Keeping in mind that $p^{ij}$ is a tensor density of weight one and using the properties of the spherical adapted coordinates we can write the constraint equations in the form
\begin{subequations}
\begin{align}
\partial_{r}p_{\phantom{r}A}^{r}\ &=\ -\partial_{B}p_{\phantom{B}A}^{B}+\Gamma_{\phantom{C}AB}^{C}p_{\phantom{B}C}^{B}-\frac{1}{2}C^{\rm matt}_{A}(\phi_\alpha,\pi^\alpha), \\
\partial_{r}p_{\phantom{r}r}^{r}\ &=\ \frac{1}{2}q_{AB,r}p^{AB}-\partial_{A}p_{\phantom{A}r}^{A}-\frac{1}{2}C^{\rm matt}_{r}(\phi_\alpha,\pi^\alpha).
\end{align}
\end{subequations}
Geometrically, we can understand those equations in terms of the family of 2-surfaces $r\ =\ {\rm const}$, and induced on each of them $r$-dependent: 2-metric $q_{AB}$, its covariant 2-derivative ${\cal D}$, a 2-tensor density $p^{AB}$, a 2-vector density $p^A{}_r$. The Christoffel symbols $\Gamma^C{}_{AB}$ on the right hand side of the first equation correspond to the covariant derivative ${\cal D}$, also $\partial_{A}p_{\phantom{A}r}^{A}={\cal D}_{A}p_{\phantom{A}r}^{A}$ and  $p_{\phantom{A}r}^{A}=q^{AB}p_{\phantom{r}A}^{r}$. In terms of those structures, the equations read
\begin{subequations}\label{theequationboth}
\begin{align}
\partial_{r}p_{\phantom{r}A}^{r}\ &=\ -{\cal D}_{B}p_{\phantom{B}A}^{B} -\frac{1}{2}C^{\rm matt}_{A}(\phi_\alpha,\pi^\alpha)\label{theequation1},\\
\partial_{r}p_{\phantom{r}r}^{r}\ &=\ \frac{1}{2}q_{AB,r}p^{AB}-{\cal D}_{A}p_{\phantom{A}r}^{A}-\frac{1}{2}C^{\rm matt}_{r}(\phi_\alpha,\pi^\alpha).\label{theequation2}
\end{align}
\end{subequations}

To construct a solution of (\ref{theequationboth}) functions $q_{AB}$, $p^{AB}$, $\phi_\alpha$ and $\pi^\alpha$ can be chosen and fixed freely, modulo suitable consistency conditions which ensure that they define a continuous and differentiable, respectively, 3-metric $q_{ij}$, 3-tensor density $p^{ij}$, and fields $\phi_\alpha$ and $\pi^\alpha$ at the point $\sigma_0$. We will see that $p^{rr}$ and $p^{rA}$ are determined by the above equations given that data.

First, let us ignore the consistency conditions at $\sigma_0$, that is at $r=0$. The first equation (\ref{theequation1}) is a family of ordinary differential equations along the rays $\theta={\rm const}$ parametrised by $r$. The unknown is $p^r{}_A(r,\theta)$, while the right hand side is given by the fixed $q_{AB}(r,\theta)$, $p^{AB}(r,\theta), \phi_\alpha(r,\theta)$ and $\pi^\alpha(r,\theta)$. Therefore, given the right hand side, the equation (\ref{theequation1}) determines  $p^r{}_A(r,\theta)$ modulo an unknown initial value $p^r{}_A(0,\theta)$. The second equation (\ref{theequation2}) is, again, a family of ordinary differential equations along the rays $\theta={\rm const}$. In this equation, the unknown is $p^r{}_r(r,\theta)$. The right hand side is given by a solution $p^r{}_A(r,\theta)$ of the first equation and by the fixed $q_{AB}(r,\theta)$, $p^{AB}(r,\theta), \phi_\alpha(r,\theta)$ and $\pi^{\alpha}(r,\theta)$. Therefore, the second equation determines  $p^r{}_r(r,\theta)$ modulo a free initial value $p^r{}_r(0,\theta)$. 

Let us turn to the consistency conditions at $r=0$. To spell them out for the tensor density $p^{ij}$ let us use the Cartesian adapted coordinates $(x^I)$ which in contrast to the spherical coordinates $(y^a)$ are well-defined in an entire neighborhood of $\sigma_0$ including $\sigma_0$ itself. A necessary and sufficient condition for $p^{ab}$ to be extendable to a tensor density continuous and $n$-times differentiable at $\sigma_0$ is that there exist functions $p^{IJ}$ continuous and differentiable $n$-times at $\sigma_0$ such that
\begin{equation}\label{ijab}
p^{ab}\ =\ r^2\Omega(\theta) \frac{\partial y^a}{\partial x^I}\frac{\partial y^b}{\partial x^J}p^{IJ},
\end{equation}
where 
\begin{equation}
\Omega(\theta)\ =\ \frac{1}{r^2}\det\left(\frac{\partial x^I}{\partial y^a}\right)
\end{equation}
is a function of the angles only. The functions $p^{IJ}$ are just the components of the tensor density $p$ in the coordinates $(x^I)$. Therefore, for continuity, the functions $p^{AB}$ we choose and fix in equation (\ref{theequationboth}) have to satisfy a condition that there are constants $p^{IJ}_0$ such that
\begin{equation}\label{piAB0}
p^{AB}(0,\theta)\ =\ s^{AB}_{IJ}(\theta)p^{IJ}_0,
\end{equation}
where the functions $s^{AB}_{IJ}$ are defined to be
\begin{equation}
s^{AB}_{IJ}\ =\ r^2\Omega(\theta) \frac{\partial y^A}{\partial x^I}\frac{\partial y^B}{\partial x^J}
\end{equation}
and indeed, each of them is independent of $r$.

Equation (\ref{ijab}) implies the continuity consistency conditions also for $p^{rA}$ and $p^{rr}$ at $\sigma_0$. As a consequence, the initial values in the equations (\ref{theequationboth}) are not any longer free:
\begin{equation}
p^{rA}(0,\theta)\ =\ 0\ =\ p^{rr}(0,\theta).
\end{equation}
One could be concerned, that with this condition a part of the data $p^{IJ}_0$ is lost. However, given $p^{AB}$ (\ref{piAB0}), all the numbers $p^{IJ}_0$ are determined due to the dependence of the function $s^{AB}_{IJ}$ on $\theta$. In conclusion, the components $p^{rr}$ and $p^{rA}$ of the gravitational momentum are completely determined by equations (\ref{theequationboth}) and by the remaining components of the momentum, and the remaining fields. More specifically, each value $p^{rA}(r,\theta)$ and, respectively, $p^{rr}(r,\theta)$ is a function of the values $p^{AB}(r',\theta)$, $q_{AB}(r',\theta)$, $\phi_\alpha(r',\theta)$ and $\pi^\alpha(r',\theta)$ taken when $r'$ ranges over the interval $(0,\ r]$. 

Finally, this relation passes unchaged to the observables $P^{rr}(r,\theta),P^{rA}(r,\theta^A)$. Namely, they can be expressed by the observables $Q_{AB}(r',\theta)$, $P^{AB}(r',\theta)$, $\Phi_\alpha(r',\theta)$ and $\Pi^\alpha(r',\theta)$ using the exact same formulas with which $p^{rr}$ and $p^{rA}$ were determined if only all the fields are replaced by the corresponding observables.

\section{Summary}

In the presented paper we have proposed and studied a new scheme of deparametrization of general relativistic systems.
Our aim, the general way we want to implement it, the starting point and the notation, have been explained in 
 sections \ref{Introduction} and \ref{Idea}. Here, we outline and interpret the results established in this paper.
     
On the phase space $\Gamma$ formed by canonical ADM data $(q,p,\phi_\alpha,\pi^\alpha)$ we introduced observables 
(see section \ref{invobsinadaptedcoordinates}). The observables are labelled by using the set of labels $\mathbb{R}^3$ 
which comes with the natural Cartesian coordinates $\mathbb{R}^3\ni x = (x^1,x^2,x^3) = (x^I)$. 
We also use thereon the natural spherical coordinates $(r,\theta^1,\theta^2)= (r,\theta)$.  The extra structure we endowed the 3-manifold $\Sigma$ (on which the ADM data is defined) with is: 
  \begin{itemize}
  \item a point $\sigma_0 \in \Sigma$;  
  \item a family of frames tangent to $\Sigma$ in $\sigma_0$, such that the transition matrix between each two of them 
  is lower-triangular and has positive entries on the diagonal;
  \end{itemize}
 one can think of that structure as of an observer, observers frames and observers coordinates.   
 For every metric tensor $q$ defined on $\Sigma$, exactly one of the observers frames, say $e_I$, is orthonormal. 
 Given $q$, we use this frame to define a map
\begin{equation}\label{summexp}
\mathbb{R}^3\ni(x^I)\ \mapsto \exp_{\sigma_0} (x^I e_I)\in \Sigma ,
\end{equation}
defined by the geodesic curves in $\Sigma$ beginning at $\sigma_0$. The observables introduced in this paper are the components of the fields on  $\mathbb{R}^3$ obtained by the pullback of the ADM data fields $q,p,\phi_\alpha, \pi^\alpha$ with the map (\ref{summexp}). Written in terms of the Cartesian coordinates $(x^I)$ they constitute the observables
\begin{equation}\label{summobs}
Q_{IJ}(x), P^{IJ}(x), \Phi_\alpha(x), \Pi^\alpha(x) :\Gamma \rightarrow \mathbb{R}
\end{equation}   
and written in terms of the spherical coordinates in the observers space $\mathbb{R}^3$ they are
\begin{equation}
Q_{AB}(r,\theta), P^{AB}(r,\theta), P^{rA}(r,\theta), P^{rr}(r,\theta), \Phi_\alpha(r,\theta), \Pi^{\alpha}(r,\theta) :\Gamma \rightarrow \mathbb{R}^3.
\end{equation}
The map (\ref{summexp}) is defined for arbitrary $x$. Therefore, for the metric tensor field $q$ and a scalar field
$\phi_\alpha$ the pullbacks $Q$, and respectively, $\Phi_\alpha$, are defined on all the observers coordinate space $\mathbb{R}^3$.
So are the corresponding observables $Q_{IJ}(x)$ and $\Phi_\alpha(x)$.  However, given $q$,  the values of the observables $Q_{IJ}(x)$ may vanish for some $x$, hence  
\begin{equation}
Q_{x} = Q_{IJ}(x)dx^Idx^J
\end{equation}
may become degenerate as a metric tensor in some points $x\in \mathbb{R}^3$. On the other hand, the fields carrying the contravariant indices like the momentum $p$ canonically conjugate to the metric tensor $q$, or any matter covector field, are defined well only on the $q$ dependent open neighborhood of $(0,0,0)\in \mathbb{R}^3$ such that the map (\ref{summexp}) maps it diffeomorphically into $\Sigma$. Therefore, we restricted our considerations to that neighborhood. The consequence for the corresponding observables is that given $x\in \mathbb{R}^3$,  the corresponding 
observables $P^{IJ}(x)$ etc, are defined only in an open neighborhood in $\Gamma$ consisting of the points $(q,p,\phi_\alpha,\pi^\alpha)$ such that the $q$-dependent map (\ref{summexp}) maps diffeomorphically a neighborhood of $x$ in $\mathbb{R}^3$ into $\Sigma$.
       
The observables are invariant with respect to the induced action in $\Gamma$ of those diffeomorphisms of $\Sigma$ 
which preserve the chosen family of frames tangent to $\Sigma$ at $\sigma_0$. This subgroup of the diffeomorphism group 
is denoted throughout the work by Diff$_{\text{obs}}$. 

We have also characterized the remaining, residual diffeomorphisms of $\Sigma$, whose induced action in $\Gamma$ 
does not leave our observables invariant. Their generators in $\Sigma$ are given by equation (\ref{resultingDelta}). For every metric tensor $q$, they form a 6 dimensional family of residual vector fields defined in the suitable neighborhood of $\sigma_0\in\Sigma$ (and extended arbitrarily to the entire $\Sigma$). A general vector field in $\Sigma$ was decomposed in sections \ref{ageneralconsideration} into a generator of Diff$_{\text{obs}}$ and one of the residual vector fields (\ref{resultingDelta}).     

The variations $\int_\Sigma d^3 \sigma\ w_{ab}(\sigma) \frac{\delta}{\delta q_{ab}(\sigma)}$ of our observables (\ref{summobs}) 
are the key technical element in the calculation of their Poisson bracket. To every variation $w_{ab}$ we assigned a vector
field $\vec\Lambda$ (given by \ref{NinPois}) in observers coordinate space $\mathbb{R}^3$. The variation
$\int_\Sigma d^3\sigma\ w_{ab}(\sigma) \frac{\delta}{\delta q_{ab}(\sigma)}F(x)$  (\ref{thevariation}) of each of our 
observables $F\in\{Q_{IJ}, P^{IJ},\Phi_\alpha,\Pi^\alpha$\} is expressed by the Lie derivative ${\cal L}_{\vec\Lambda}F(x)$
in the observers coordinate space.  

Another special property of the assignment $w\mapsto\vec\Lambda$ is the dependence of $\vec\Lambda(x)\in T_x \mathbb{R}^3$  
on $w_{ab}(\sigma_0)$. The point $\sigma_0$ contributes with a measure proportional to the Dirac delta (namely $d^3x' \delta(0,x')$) and / or its derivative depending on the type of the observed field.  
Other points of the geodesic curve connecting the point $\sigma_{(x)}$ (the image of (\ref{summexp})) with the origin $\sigma_0$ contribute to the vector  $\vec\Lambda(x)$ with a measure proportional to $d^2\theta \delta(\theta-\theta(\sigma))$ (and / or derivatives), where $\theta = (\theta^1(\sigma_{(x)}),\theta^2(\sigma_{(x)})$ are the values of the spherical coordinates. This property has important consequences for the Poisson bracket between the observables.

The complete understanding of the variations of the observables allowed us to calculate their Poisson algebra in section \ref{poissonbracketofobs}. The Poisson bracket between the observables $Q_{IJ}(x),\Phi_{\alpha}(x),\Pi^{\alpha}(x)$ is canonical 
(\ref{PBcan}), in the sense that $\Phi_\alpha(x)$ and $\Pi^{\alpha}(x)$ are canonically conjugate to each other and both Poisson commute with all the $Q_{IJ}(x)$, while $Q_{IJ}(x)$ Poisson commute among each other.
On the other hand, the Poisson bracket  $\{F(x),\ \int d^3x' w_{IJ}(x')P^{IJ}(x')\}$ carries all the nontrivial structure of our observables.
For $F=Q_{IJ},\Phi_\alpha,\Pi^\alpha$ it is given by (\ref{PBPIJ}), while for $F=P^{IJ}$ it is provided by (\ref{PIJPIJ}). In both cases the Poisson bracket is expressed by the Lie derivative of the observables with respect to the vector field defined on the observers coordinate space, therefore the result is given in terms of the observables. For $x\neq0$ it is convenient to use the observers spherical coordinates. Then, the observables $Q_{AB}(r,\theta),P^{AB}(r,\theta),\Phi_\alpha(r,\theta), \Pi^{\alpha}(r,\theta)$ have canonical Poisson brackets provided we consider only functionals $\int drd^2\theta\ f_{AB}(r,\theta)P^{AB}(r,\theta)$ with a smearing function whose support does not contain $\sigma_0$. The full Poisson bracket, though, contains the contribution 
from the point $(0,0,0)$ (\ref{PBPAB}).

The observables $\Phi_\alpha(x)$, $\Pi^\alpha(x)$ are defined for arbitrary bosonic matter fields and the index $\alpha$
may give them arbitrary tensor character. For simplicity, in many places of our paper we restrict ourselves to scalar matter fields. For instance, the explicit formulae (\ref{thevariation}) for the variation of the matter fields and, respectively, (\ref{PBPIJ}) for their Poisson brackets with $P^{IJ}$ assume the scalar case. However, a generalization of the Lie derivatives ${\cal L}_{\vec\Lambda} \Phi_{\alpha}$ and ${\cal L}_{\vec\Lambda} \Pi^{\alpha}$ to arbitrary vector or tensor matter fields gives formulae true for those more general cases of matter.    
 
The framework introduced in this paper can be applied to canonical theories of fields $(q,p,\phi_\alpha,\pi^\alpha)$ constrained by the vector constraints. For those theories the action of the diffeomorphisms of $\Sigma$ induced in $\Gamma$ coincides with the gauge transformations. The two examples we refer to in section \ref{dynamo} are the Rovelli-Smolin model of general relativistic theory deparametrised by a massless scalar field and the Brown-Kucha$\check{\rm r}$ model of a general relativistic theory deparametrised by
non-rotating dust. We applied the results of the previous sections to derive the dynamics of the observables (\ref{summobs}) and to write it again in terms of the observables. This result is contained in equation (\ref{dynobs}). The contribution to the dynamics of any observable $G(x)$, comes from the other observables $G(x')$ at $x'=(0,0,0)$ and at $x$ with the point measure, and also integrated along the line segment from $(0,0,0)$ to $x$. The Hamiltonian globally defined in terms of the fields $(q,p,\phi_\alpha,\pi^\alpha)$ can be replaced by an equivalent local Hamiltonian defined in terms of the observables (\ref{summobs}). 

Finally, in section \ref{sekcjarozwdiff} we address the vector constraints. On the vector constraint surface we express the observables $P^{rr}(r,\theta)$ and $P^{rA}(r,\theta)$ by $Q_{AB}(r',\theta), P^{AB}(r',\theta)$, $\Phi_\alpha(r',\theta)$ and $\Pi^\alpha(r',\theta)$. The latter observables are free modulo the assumption that they are the components in the spherical coordinates of corresponding fields $Q_{IJ}$, $P^{IJ}$, $\Phi_\alpha$ and $\Pi^\alpha$ not singular in $(0,0,0)$.

The results of this work open the door to the reduced degrees of freedom formulation of our framework, obtained by restricting our observables to the vector constraint surface in $\Gamma$. The reason for which in this paper we were working in the whole kinematical phase space $\Gamma$ was to control all the subtleties following from the existence of the residual gauge transformations our observables are sensitive to. Now we have reached the point at which the passage to the reduced phase space is possible.

\section{Acknowledgements}

This work was partially supported by the grant of Polish Narodowe Centrum Nauki nr 2011/02/A/ST2/00300 and by the grant of Polish Narodowe Centrum Nauki nr 2013/09/N/ ST2/04299.

\end{document}